\documentclass[aip,pop,superscriptaddress,english,graphicx,floatfix]{revtex4-1}

\usepackage[T1]{fontenc}
\usepackage[latin9,utf8]{inputenc}
\usepackage{calc}
\usepackage{units}
\usepackage{verbatim}

\usepackage{epsfig}
\usepackage{subfigure}
\usepackage{wrapfig}
\usepackage{graphicx}

\usepackage{dcolumn}
\usepackage{multirow}
\usepackage{dcolumn}
\newcolumntype{d}[1]{D{.}{.}{#1}}
\usepackage{booktabs} 

\usepackage{amsmath}
\usepackage{amssymb}
\usepackage{amsthm}
\usepackage{amsfonts}
\usepackage{amstext}
\usepackage{mathtools}
\usepackage{array} 
\usepackage{cancel}
\usepackage{algorithm}
\usepackage{algorithmicx}
\usepackage{algpseudocode}
\usepackage{bm}
\usepackage[mathlines]{lineno}
\usepackage{mathrsfs}

\usepackage[usenames,dvipsnames]{color} 
\definecolor{LightBluishGray}{rgb}{0.95,0.95,1.0}
\definecolor{BlueSky}{RGB}{0,127,255}
\definecolor{shade50}{rgb}{0.5,0.5,0.5}
\definecolor{shade20}{rgb}{0.8,0.8,0.8}
\definecolor{shade10}{rgb}{0.9,0.9,0.9}
\definecolor{shade05}{rgb}{0.95,0.95,0.95}
\definecolor{red50}{rgb}{0.5,1.0,1.0}
\definecolor{blue50}{rgb}{1.0,1.0,0.5}
\usepackage{soul}
\sethlcolor{yellow}

\usepackage[usenames,dvipsnames]{pstricks}
\usepackage{pst-grad} 
\usepackage{pst-plot} 

\usepackage{paralist} 
\usepackage{titlesec}
\usepackage{enumitem}
\setlist{nolistsep}
\usepackage{fancyhdr}
\usepackage[title,titletoc]{appendix}
\usepackage{changepage}
\usepackage{pdfpages}

\usepackage[margin=1.0in]{geometry}
\titlespacing\section{0pt}{6pt plus 2pt minus 2pt}{0pt plus 2pt minus 2pt}
\titlespacing\subsection{0pt}{6pt plus 2pt minus 2pt}{0pt plus 2pt minus 2pt}
\titlespacing\subsubsection{0pt}{6pt plus 2pt minus 2pt}{0pt plus 2pt minus 2pt}
\setlist[itemize]{topsep=0pt}
\setlist[enumerate]{topsep=0pt}

\allowdisplaybreaks[2]

\usepackage{babel}
\usepackage{natbib}
\usepackage{hyperref}
\hypersetup{
  colorlinks   = true, 
  urlcolor     = blue, 
  linkcolor    = blue, 
  citecolor   = red     
}
\usepackage{cleveref}
\crefname{section}{part}{parts}
\crefname{section}{section}{sections}

\newenvironment{frcseries}{\fontfamily{frc}\selectfont}{}

\DeclareFontFamily{OT1}{pzc}{}
\DeclareFontShape{OT1}{pzc}{m}{it}{<-> s * [1.10] pzcmi7t}{}
\DeclareMathAlphabet{\mathpzc}{OT1}{pzc}{m}{it}

\newcommand{\Ncal}{\ensuremath{\mathcal{N}}}

\newcommand{\Zcal}{\ensuremath{\mathcal{Z}}}

\newcommand{\avrg}[1]{\ensuremath{\langle{#1}\rangle}}
\newcommand{\abs}[1]{\ensuremath{\lvert{#1}\rvert}}
\newcommand{\bvec}[1]{\ensuremath{\mathbf{#1}}}
\newcommand{\tvec}[1]{\ensuremath{\tilde{\bvec{#1}}}}

\renewcommand{\bar}[1]{\overline{#1}}
\newcommand{\cart}[1]{\ensuremath{\hat{\bvec{#1}}}}
\newcommand{\avom}[1]{\avrg{#1}_{_{\Omega'} } }

\begin{document}

\title{Modeling of Inelastic Collisions in a Multifluid Plasma:\\ Excitation and Deexcitation\footnote{Distribution A. Approved for public release; distribution unlimited}} 

\author{Hai P. Le}
\thanks{Corresponding author}
\email{hai.le@ucla.edu}
\affiliation{Department of Mathematics, University of California, Los Angeles, California 90095}
\author{Jean-Luc Cambier}
\email{jean\_luc.cambier@us.af.mil}
\affiliation{Air Force Research Laboratory, Edwards AFB, California 93524}

\date{\today}

\begin{abstract}
We describe here a model for inelastic collisions for electronic excitation and deexcitation processes in a general, multifluid plasma. The model is derived from kinetic theory, and applicable to any mixture and mass ratio. The principle of detailed balance is strictly enforced, and the model is consistent with all asymptotic limits. The results are verified with direct Monte Carlo calculations, and various numerical tests are conducted for the case of an electron-hydrogen two-fluid system, using a generic, semi-classical model of collision cross sections. We find that in some cases, the contribution of inelastic collisions to the momentum and thermal resistance coefficients is not negligible, in contrast to the assumptions of current multifluid models. This fundamental model is also applied to ionization and recombination processes, the studies on which are currently underway.
\end{abstract}

\maketitle

\section{Introduction}
Modeling of nonequilibrium processes in a low-temperature partially ionized plasma is of particular interest to a wide range of technical fields such as gas discharge, electric propulsion, spectroscopic and laser diagnostics, and material science~\cite{raizer_gas_1997,lieberman_principles_2005,jahn_physics_2006}. 
The complexity of the model is largely due to the characterization of various collisional and radiative processes, occurring at a wide range of spatial and temporal scales \cite{biberman_kinetics_1987,zeldovich_physics_2002}. Although the fundamental physical processes may be individually known, it is not always clear how their combination affects the overall operation, or at what level of detail this process needs to be modeled. 
The current state of the art for modeling detailed chemical kinetics of a low temperature plasma is the collisional-radiative (CR) model, first proposed by Bates et al. in 1962 \cite{bates_recombination_1962,bates_recombination_1962-1}. CR models are now commonly used in studies of plasma discharge, plasma-assisted combustion, and hypersonics \cite{vlcek_collisional-radiative_1989,poggie_numerical_2013,panesi_electronic_2011,kapper_ionizing_2011,kapper_ionizing_2011-1,capitelli_fundamental_2013}. 
The advantage of a CR model is two-fold. First, strong deviations from equilibrium of the internal states can be captured accurately when CR models are employed. In addition, \textit{ab initio} cross section data can be directly incorporated in the CR model, leading to a very accurate prediction of the thermochemical kinetics of the system. 

There are, however, two issues with this modeling approach. The first arises from the complexity of the physical processes needed to be captured in the model. The required level of detail of the CR model is typically not known a priori and is possibly changing in a dynamical fashion as the system evolves in time. This can be resolved by coarse-graining techniques, which reduce the complexity of the kinetics to avoid solving a large system of equations \cite{le_complexity_2013,panesi_collisional_2013,guy_consistent_2015}. 
The second issue comes from translational nonequilibrium, often found in a discharge, where we have both a bulk plasma (continuum) and a highly energetic component (kinetic), e.g., electrons emitted from the cathode. A proper treatment of this energetic beam-like component requires extending the solution of the CR kinetics to the so-called non-Maxwelllian regime \cite{vahedi_monte_1995,hagelaar_solving_2005,yan_analysis_2015}. These simulations are typically very expensive and therefore limited to zero- or one-dimensional systems. In addition, space charge effects within the bulk plasma can also become important, requiring further separation, i.e., ions and electrons.
A natural solution to this problem is to use a hybrid approach, decomposing the system into a continuum and a kinetic component \cite{beouf_pseudospark_1991}. The continuum component can be solved by fluid equations, and the kinetic component by (for example) a particle method. Although the idea seems quite intuitive, proper treatment of the coupling between the fluid and kinetic components is highly non-trivial. There appears to be no unique and consistent coupling methodology, and the choice is highly problem-dependent. This coupling issue is currently being addressed by many researchers and is outside of the scope of the current work.

We are concerned here with an alternative approach, the so-called multifluid model, which decomposes the plasma into several fluid components. For example, in a discharge configuration, one could have 4 different fluids, namely the neutrals, ions, bulk electrons, and the energetic electrons. The only required assumption is that collisions among particles within the same fluid are sufficiently fast to maintain a Maxwellian distribution. The validity of this assumption is not always well known. Nevertheless this approach is attractive, since it is much faster than a fully kinetic solver, and unambiguous since at the same time, the approach relies on kinetic theory for the treatment of coupling terms between different fluids. Furthermore, one can rely on fast implicit methods to solve these fluid equations and examine long time behavior of the system; this offers a big advantage over kinetic simulations, which are only suitable for problems with short time scales. Multifluid models are commonly used in simulations of astrophysical plasmas (see for example \cite{zaqarashvili_magnetohydrodynamic_2011,leake_multi-fluid_2012}).

The most classical work in multifluid plasma modeling is due to Braginskii \cite{braginskii_transport_1965}, who derived fluid equations for a fully ionized plasma, using a Chapman-Enskog closure. Braginskii's work has been successively refined by several authors, with particular emphases on improving the transport coefficients and/or including interaction with neutral species \cite{decoster_modeling_1998,zhdanov_transport_2002,meier_general_2012,khomenko_fluid_2014}. Burgers, on the other hand, presents a rather general framework for the modeling of elastic collisions \cite{burgers_flow_1969}. These include both neutral collisions and charged particle collisions; the methodology is applicable for a general system of moment equations beyond the standard five-moment model. Burgers also introduces a simplified model for reactive collisions using a Bhatnagar-Gross-Krook (BGK) collision operator. 

In the current work, we present a self-consistent model for inelastic collisions within the multifluid framework. The model is derived from kinetic theory and obeys the principle of detailed balance (DB), which we show to be an essential property to ensure that the system approaches the correct equilibrium limit. We focus on characterizing the exchange source terms due to collision, namely mass, momentum and energy exchanges (the hydrodynamics and transport fluxes can be added following~\cite{braginskii_transport_1965} when considering a non-uniform plasma). We will show that in most cases none of these terms can be neglected, and they have complex dependencies to microscopic quantities of the interaction, e.g., multiply-differentiated cross sections. 
We present the general description of the collision kinematics and derive the exchange terms for the case of excitation and deexcitation processes. Although we are mostly interested here in electron-impact collisions and atomic transitions, we keep the formulation as general as possible, such that the application to other species and chemistry (e.g. proton-impact, molecular vibrational transitions, charge-exchange, etc.) is a straight-forward extension.  The case of ionization and recombination, and other three-body processes, is currently under examination, using the same basic formulation presented here. 

The rest of the paper is organized as follows. The derivation of the exchange source terms is given in Sec. \ref{sec:rates}, by first  introducing the description of the transfer integral, and then presenting the derivation of the exchange rates in the following subsections. In Sec. \ref{sec:numerix}, we show the numerical evaluation of the multifluid rates, verify the results with Monte Carlo calculations, and perform zero-dimensional calculations utilizing the multifluid rates. Finally, conclusions and a summary of the present findings are given in Sec. \ref{sec:conclusion}. We also provide several appendices to elaborate on the derivation of the exchange source terms and the description of the numerical simulation. 

\section{Rate Derivation}\label{sec:rates}

\subsection{Transfer integral}\label{sec:transfer}

Let us consider an inelastic collision between two particles $s$ and $t$, such that the particle $t$ changes its internal state. The particles $s$ and $t$ are respectively the scattered and target in the laboratory frame of reference (LAB). The former will be identified as the electron and the target as the atom, but we will keep the general $s,t$ notation until explicit assumptions and approximations are made, such as neglecting terms of the order of the mass ratio $m_s/m_t$ for final expressions. Following Appendix~\ref{app:A}, the initial velocities are $\bvec{v}_s, \bvec{v}_t$, where $\bvec{v}\!=\!\bvec{u}\!+\!\bvec{c}$ and $\bvec{u}$ is the fluid mean velocity in the LAB frame, and post-collision values are indicated by a prime, i.e.:
\begin{align}\label{eq:ti1}
s(\bvec{v}_s) + t(\bvec{v}_t) \rightarrow s' (\bvec{v'}_s) + t'(\bvec{v'}_t)
\end{align} 

We make here two assumptions: 1) the collision produces only two particles, which may or may not belong to the same fluids as the initial reactants, and 2) the masses of individual particles are the same before and after the collision, e.g.  $m'_{s}\equiv m_s$, such that mass conservation is automatically obtained. Both of these will be revisited in a follow-on paper, dealing with ionization and recombination. 
Defining the energy transfer to and from the internal modes to be represented by $\Delta\varepsilon$, we have the following energy conservation constraint on the relative velocity $\bvec{g}$ where $\bvec{g} = \bvec{v}_s - \bvec{v}_t$ (see Appendix~\ref{app:A}):
\begin{equation}\label{eq:ti2}
	\bvec{g}^2=\bvec{g'}^2+\frac{2\Delta\varepsilon}{\mu}
\end{equation}
For excitation, the transferred energy is a positive and fixed value $\Delta\varepsilon\equiv \varepsilon^*$, the energy gap between the levels, while for ionization it is a continuum of values: $\Delta\varepsilon \in \left[\varepsilon^*,\varepsilon\right]$, where $\varepsilon\!=\!\frac{1}{2}\mu\bvec{g}^2$ is the available kinetic energy in the center-of-mass (COM) frame. In the limit $\Delta\varepsilon\rightarrow 0$, the collision is elastic. We will keep the same relations for the reverse process, for which the primed variables are post-collision and non-primed refer to pre-collision, such that for deexcitation, $\Delta\varepsilon\!=\!-\varepsilon^*$.

\begin{figure}
\centering
\includegraphics[scale=1]{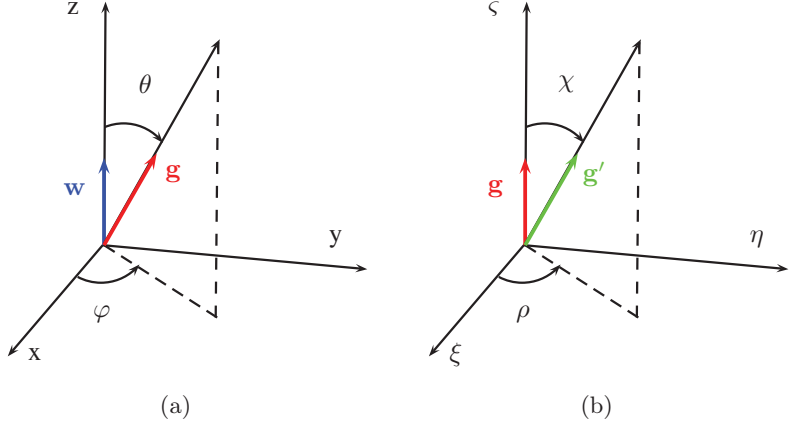}
\caption{Frame rotation and relative orientation of (a) $\bvec{w}$ and $\bvec{g}$ and (b) $\bvec{g}$ and $\bvec{g'}$. The rotation operator matrix $R(\varphi,\theta)$ (or $R(\rho,\chi)$) is defined such that $\hat{\bvec{g}} = R(\varphi,\theta)\cdot\hat{\bvec{w}}$ and $\hat{\bvec{g}}' = R(\rho,\chi)\cdot\hat{\bvec{g}}$.}
\label{fig:coord}
\end{figure}

We can then define a transfer integral of the collision operator between the two species $s$ and $t$~\cite{burgers_flow_1969}.
\begin{equation}\label{eq:ti3}
	\Psi_{st} = n_s n_t \int\! d^3\bvec{v}_s d^3\bvec{v}_t \,f_s f_t\, g \int \psi \, d\omega(\bvec{v}_s,\bvec{v}_t;\bvec{v'}_s,\bvec{v'}_t) 
\end{equation} 
where $g$ is the magnitude of the relative velocity ($g\!=\!\abs{\bvec{g}}$), $d\omega$ is the differential cross section (DCS), and $\psi$ is any moment variable exchanged during the collision. We now follow Appendix~\ref{app:B}, starting with the following transformations:
\begin{subequations}\label{eq:ti4}
\begin{align}
	\bvec{V^*} &= \bvec{V}-\bvec{U}+ \gamma \tvec{g}  &T^*= \frac{M T_sT_t}{m_sT_t\!+\!m_tT_s}\qquad\quad & a^2  = \frac{2kT^*}{M}\\
	\tvec{g} &=\bvec{g}-\bvec{w}   &\tilde{T}=\frac{m_sT_t\!+\!m_tT_s}{M}\qquad\quad &\alpha^2 = \frac{2k\tilde{T}}{\mu} \\
	\textrm{and}\quad\gamma &= \frac{\mu (T_t-T_s)}{m_sT_t\!+\!m_tT_s} 
\end{align}
\end{subequations}
where the relative mean velocity $\bvec{w} = \bvec{u}_s - \bvec{u}_t$. The product of the two Maxwellian distributions $f_s \cdot f_t$ is expressed in terms of the product of two other Maxwellians, $f_{V^*} \cdot f_{\tilde{g}}$, for the COM velocity and relative velocity respectively. 
Inserting (\ref{eq:b14}--\ref{eq:b16})  into (\ref{eq:ti3}), the transfer integral can be written as follows:
\begin{equation}\label{eq:ti5}
\Psi_{st} = n_s n_t  \frac{1}{\pi^\frac{3}{2} a^3} \int d^3 \bvec{V^*} e^{-\bvec{V}^{*2}/a^2}\cdot 
\frac{1}{\pi^\frac{3}{2} \alpha^3}\int d^3 \bvec{g} \, e^{-\tvec{g}^2/\alpha^2} g \int \psi  d\omega(\bvec{g};\bvec{g'}) 
\end{equation} 
Note that in the COM reference frame, the DCS only depends on the relative velocities, i.e., $d\omega(\bvec{v}_s,\bvec{v}_t;\bvec{v'}_s,\bvec{v'}_t) \equiv d\omega(\bvec{g};\bvec{g'})$,  and can be expressed as:
\begin{equation}\label{eq:ti6}
d\omega(\bvec{g};\bvec{g'}) \!=\!  \sigma_{st} (g,\Omega') d\Omega'
\end{equation}
where $\Omega'$ is the solid angle between the initial and final relative velocities, i.e., $d\Omega'\!=\!d\rho\,d\!\cos\chi$ with $\bvec{g}\cdot\bvec{g'}=g g'\,\cos\chi$. Without loss of generality, we can now choose a reference frame (LAB) such that the relative \emph{mean} velocity $\bvec{w}$ is aligned with the $\cart{z}$ axis, as shown in Figure~\ref{fig:coord}. Thus, the unit vectors $\hat{\bvec{g}}$, $\hat{\bvec{g}}'$ are obtained by subsequent rotations of the $(\cart{x},\cart{y},\cart{z})$ frame.
Using the abbreviated notation $c_\varphi\!\equiv\!\cos\varphi$, $s_\varphi\!\equiv\!\sin\varphi$, etc, we define this rotation operator by the matrix:
\begin{equation}\label{eq:ti7}
R(\varphi,\theta) = \left(\begin{array}{ccc} c_\varphi c_\theta & -s_\varphi & c_\varphi s_\theta \\
                                                                          s_\varphi c_\theta & c_\varphi  & s_\varphi s_\theta \\
                                                                         -s_\theta                &  0             & c_\theta      \end{array}\right)
\quad\textrm{and}\qquad
\hat{\bvec{g}}=\frac{\bvec{g}}{g} = R(\varphi,\theta)\cdot\hat{\bvec{z}} = \left(\begin{array}{c}c_\varphi s_\theta \\ s_\varphi s_\theta \\ c_\theta\end{array}\right)
\end{equation}
Similarly, the post-collision relative velocity is rotated by the angles ($\rho,\chi$), such that $\hat{\bvec{g}}' \!=\! R(\rho,\chi)\cdot\hat{\bvec{g}}$.

Using $d^3 \bvec{g} = g^2 dg d\varphi dc_\theta$, and equation (\ref{eq:ti6}), the transfer integral can be written as:
\begin{equation}\label{eq:ti8}
\begin{split}
\Psi_{st} =  \frac{n_s n_t}{\pi^{\frac{3}{2}} \alpha^3} e^{-w^2/\alpha^2}\cdot \int d^3 \bvec{V^*} f_{V^*} \cdot  \int dg \, g^3 \, e^{-g^2/\alpha^2} \cdot \\ \int d\varphi dc_{\theta} \, e^{2gwc_\theta/\alpha^2} \, \int d \rho dc_{\chi} \, \psi \sigma_{st} (g,\Omega')
\end{split}
\end{equation} 
Let us now assume that the moment variable can be expanded in terms of powers of $\bvec{V}^*$:
$$
	\psi = a + b \bvec{V}^* + c \bvec{V}^{*2} +\ldots
$$
where $a,b,c...$ are functions of the remaining velocity variables, and let us perform the integration over $\bvec{V}^*$. Note that we have:
\begin{align*}
\int d^3\bvec{V}^*f_{V^*} \equiv 1 &\qquad \int d^3\bvec{V}^*\bvec{V}^*f_{V^*} \equiv 0
\end{align*}
the latter by reasons of symmetry. Thus, as long as $\psi$ does not contain terms quadratic (or higher) in $\bvec{V}^*$, a condition satisfied throughout this work, we can eliminate the integration over $\bvec{V}^*$, keeping only the terms which are independent of $\bvec{V}^*$. Also by symmetry, the DCS $\sigma_{st}$ does not depend on the angle $\rho$, and we can write:
\begin{equation}\label{eq:ti9}
 \sigma_{st}(g,\Omega')\equiv \bar{\sigma}_{st}(g)\cdot \mathcal{G}(g,\chi)\quad\textrm{s.t.}\quad \int d\rho\ dc_\chi\, \mathcal{G}(g,\chi)\equiv 1
\end{equation}
where $\mathcal{G}$ is the angular-dependent DCS. More generally, we will define the averaging of any function $\psi$ over the scattering angles as:
\begin{equation}\label{eq:ti10}
\avom{\psi} = 2\pi\int_{-1}^{+1}\!dc_\chi \psi\ \mathcal{G}(g,\chi)
\end{equation}
A trivial integration over $\varphi$ yields:
\begin{equation}\label{eq:ti11}
\Psi_{st} = \frac{2n_sn_t}{\pi^\frac{1}{2}\alpha^3} e^{-w^2/\alpha^2} 
\int\!\!dg g^3 e^{-g^2/\alpha^2} \bar{\sigma}_{st}(g)\cdot \int_{-1}^{+1}\!\!dc_\theta e^{2gwc_\theta/\alpha^2} \avom{\psi}
\end{equation}
We now define the following, normalized energy variables,
\begin{equation}\label{eq:ti12}
z=\frac{\frac{1}{2}\mu g^2}{k\tilde{T}}\qquad  \lambda=\frac{\frac{1}{2}\mu w^2}{k\tilde{T}} \qquad z^* = \max\left(0,\frac{\Delta\varepsilon}{k\tilde{T}}\right)
\end{equation}
where $\tilde{T}$ is defined in Appendix~\ref{app:B}. Using $g^3dg\!\equiv\! 2\varepsilon d\varepsilon/\mu^2$ and a further change of variables, we finally obtain:
\begin{equation}\label{eq:ti13} 
\Psi_{st} = n_sn_t \underbrace{\left( \frac{8k\tilde{T}}{\pi\mu} \right)^\frac{1}{2} }_{\bar{g}_{\tilde{T}}} e^{-\lambda} \int_{z^*}^\infty \!dz\,z\,e^{-z}\,\bar{\sigma}_{st}(z)\,\cdot\frac{1}{2}\int_{-1}^{+1}\!\!dc_\theta e^{2\sqrt{\lambda z} c_\theta}\cdot \avom{\psi}
\end{equation}
with $\bar{g}_{\tilde{T}}$ a thermal velocity based on the average temperature $\tilde{T}$.
Note that we have left the variable $\psi$ undetermined, and since it could potentially depend on all integration variables ($z, \chi, \theta$), $\psi$ must be kept inside all integrals. We will see what simplifications can be made next, depending on which moment variables we are integrating. 

The lower limit of integration, $z^*$,  is zero for elastic collisions or for exothermic reactions. Thus, equation (\ref{eq:ti13}) is a general formula, which applies equally to excitation ($z^*>0$) and deexcitation ($z^*\equiv 0$) \footnote{Alternatively, we could have extended the integration limit to $0$ in all cases and included the threshold in the definition of the cross section $\bar{\sigma}_{st}(z)$.}. 
Let us first consider an excitation collision:
\begin{equation*}
s (\bvec{v}_s) + t (E_\ell,\bvec{v}_t) \rightarrow s (\bvec{v'}_s) + t (E_u,\bvec{v'}_t)
\end{equation*}
where $\ell$ and $u$ denote the lower and upper energy states, respectively. 
We made here the assumption that both states $(\ell,u)$ belong to the same fluid, so that the particle indices $(s,t)$ are kept the same, but this is not necessary.
From eq. (\ref{eq:ti2}), energy conservation implies $\Delta \varepsilon \!=\! E_u\! -\! E_\ell > 0$. We can then define normalized energy variables for this case, $x, x'$ and $x^*$:
\begin{equation}\label{eq:ti14}
\quad x^*=\frac{\varepsilon^*}{k\tilde{T}} > 0; \qquad x'\equiv x-x^*>0
\end{equation}
Note that $x$ $(x')$ is the normalized kinetic energy of the initial (final) products of excitation respectively, and that $x^*$ is the normalized energy threshold, always a positive quantity.
For excitation, we can use (\ref{eq:ti13}) with the following identifications:
\begin{equation}\label{eq:ti15}
	z\equiv x, \quad z^*\equiv x^*,\quad z'\equiv x',\quad \textrm{and}\  \quad n_t \equiv n_\ell
\end{equation}

For a deexcitation collision, we have the reverse ($u \rightarrow \ell$), i.e. $\Delta \varepsilon < 0$:
\begin{equation}
\label{eq:dex_col}
s (\bvec{v}_s) + t (E_u,\bvec{v}_t) \rightarrow s (\bvec{v'}_s) + t (E_\ell,\bvec{v'}_t)
\end{equation}
Equation (\ref{eq:ti13}) is again still valid, if we now make the following identifications:
\begin{equation}\label{eq:ti16}
 	z\equiv x', \quad z^*\equiv 0, \quad z'\equiv x, \quad\textrm{and} \quad n_t \equiv n_u 
\end{equation}
Therefore, in all cases the variable $x$ always refers to the larger kinetic energy (before excitation or after deexcitation) and $x'$ refers to the smaller value (after excitation or before deexcitation). We can therefore define two cases of (\ref{eq:ti13}):
\begin{subequations}\label{eq:ti17} 
\begin{align}
\Psi^\uparrow_{s\ell} &= n_sn_\ell \bar{g}_{\tilde{T}} e^{-\lambda} \int_{x^*}^\infty \!dx\,x\,e^{-x}\,\bar{\sigma}^\uparrow_{s\ell}(x)\,\cdot \frac{1}{2}\int_{-1}^{+1}\!\!dc_\theta e^{2\sqrt{\lambda x} c_\theta}\cdot \avom{\psi} \\
\Psi^\downarrow_{su} &= n_sn_u \bar{g}_{\tilde{T}} e^{-\lambda} \int_{0}^\infty \!dx'\,x'\,e^{-x'}\,\bar{\sigma}^\downarrow_{su}(x')\,\cdot \frac{1}{2}\int_{-1}^{+1}\!\!dc_\theta e^{2\sqrt{\lambda x'} c_\theta}\cdot \avom{\psi} 
\end{align}
\end{subequations}
where the superscripts $\uparrow,\downarrow$ indicate excitation and deexcitation respectively (note the change of subscript from $st$ to $s\ell$ for excitation, and $su$ for deexcitation). It is worth pointing out that the averaging over the scattering angle, i.e, $\langle \psi \rangle_{\Omega'}$, has to be done with the corresponding angular-dependent DCS $\mathcal{G}$, e.g., $\langle \psi \rangle_{\Omega '} = 2\pi \int c_\chi \psi \mathcal{G}^\uparrow_{s\ell}$ for excitation. However, from time reversal we have $ \mathcal{G}^\uparrow_{s\ell} \!=\! \mathcal{G}^\downarrow_{su}$, so for simplicity, we do not differentiate $\langle \psi \rangle$ between excitation and deexcitation. 
Note that the integration over the $\theta$ angle remains to be performed. If the moment variable $\psi$ can be expanded in terms of power of $\cos\theta$, we can then define the following set of functions:
\begin{equation}\label{eq:ti18}
	\zeta^{(k)}(\xi) = \Ncal_k \int_{-1}^{+1}\!\!dy\,y^k\,e^{2\xi y}
\end{equation}
where $\Ncal_k$ is a normalizing factor. In particular, we have:
\begin{subequations}\label{eq:ti19}
\begin{align}
\zeta^{(0)}(\xi) & =\frac{1}{2}\int_{-1}^{+1}\!\!dy\,e^{2\xi y} = \frac{\sinh(2\xi)}{2\xi} &\qquad\textrm{s.t.:}\quad \lim_{\xi \to 0}\zeta^{(0)} = 1\\
\zeta^{(1)}(\xi) & =\frac{3}{4\xi}\int_{-1}^{+1}\!\!dy\,y\,e^{2\xi y} = \frac{3}{4\xi^2}\left[\cosh(2\xi)-\frac{\sinh(2\xi)}{2\xi}\right] &\qquad\textrm{s.t.:}\quad \lim_{\xi \to 0}\zeta^{(1)} = 1
\end{align}
\end{subequations}

In the CR model, each internal state is treated as a pseudo-species, so the rate of change in number density for each state ($n_\ell,n_u$) is taken into account separately. We can now examine the specific form taken by the transfer integral, according to the chosen moment variable, starting from (\ref{eq:ti8}), (\ref{eq:ti13}), or (\ref{eq:ti17}).

\subsection{Zero$^{th}$-order moment: number density}
\label{sec:zero-mom}

The rate of change of the number density due to an inelastic collision of type (\ref{eq:ti1}) can be obtained by setting $\psi\!\equiv\!1$ in (\ref{eq:ti13}), so the average over all the scattering angle is trivially removed:
\begin{equation}\label{eq:mf0_1} 
\Gamma_{st} = n_sn_t \bar{g}_{\tilde{T}} e^{-\lambda} \int_{z^*}^\infty \!dz\,z\,e^{-z}\,\bar{\sigma}_{st}(z)\,\cdot\frac{1}{2}\int_{-1}^{+1}\!\!dc_\theta e^{2\sqrt{\lambda z} c_\theta}
\end{equation}
The integration over $dc_\theta$ yields the function $\zeta^{(0)}$ defined in eq. (\ref{eq:ti19}).
We can now express the rates for transitions between two atomic levels $\ell,u$, by making the appropriate substitutions for the energy variables.
For the case of an excitation ($\ell\rightarrow u$), and according to eq. (\ref{eq:ti15}), we define the variable $x$ as the normalized kinetic energy of the reactants ($s,t$) in the COM frame, prior to the collision: therefore in this case, $z \equiv x$, $z^* \equiv x^* > 0$ and $n_t \equiv n_\ell$. Thus,
\begin{equation}\label{eq:mf0_2a}
\Gamma_{s\ell}^{\uparrow} = n_s n_\ell \bar{g}_{\tilde{T}} e^{-\lambda} \int_{x^*}^\infty dx\,x\,e^{-x}\,\bar{\sigma}^{\uparrow}_{s\ell}(x)\zeta^{(0)}(\sqrt{\lambda x})
\end{equation}
For deexcitation ($u\rightarrow\ell$), the rate of change of number density follows from (\ref{eq:ti16}): 
\begin{equation}\label{eq:mf0_2b}
\Gamma_{su}^{\downarrow} = n_s n_u \bar{g}_{\tilde{T}} e^{-\lambda} \int_{0}^\infty dx'\,x'\,e^{-x'}\,\bar{\sigma}^{\downarrow}_{su}(x')\zeta^{(0)}(\sqrt{\lambda x'})
\end{equation}
Both of these quantities are positive, hence the resultant rates equations are:
\begin{equation*}
\frac{dn_\ell}{dt} = -\Gamma_{s\ell}^{\uparrow} = -\frac{dn_u}{dt} \qquad\textrm{and}\qquad \frac{dn_\ell}{dt} = +\Gamma_{su}^{\downarrow} = -\frac{dn_u}{dt}
\end{equation*}

In the case of electron-impact processes ($s\equiv e$), we can neglect terms of order $m_e/M$, and for an atomic transition between levels $\ell\rightarrow u$, we obtain:
\begin{equation}\label{eq:mf0_3}
\Gamma^\uparrow_{e\ell}= n_en_\ell \bar{v}_e \; e^{-\lambda} \int_{x^*}^\infty\!dx\,x\,e^{-x}\,\bar{\sigma}_{e\ell}^\uparrow(x)\,\zeta^{(0)}(\sqrt{\lambda x})
\end{equation}
where $\bar{v}_e = \sqrt{ \frac{8kT_e}{\pi m_e} }$. In the limit of thermal plasma where multifluid effects are weak, i.e. $\lambda\rightarrow 0$, we obtain:
\begin{equation}\label{eq:mf0_4}
\Gamma^\uparrow_{e\ell} = n_en_\ell \bar{v}_e\int_{x^*}^\infty\!dx\,x\,e^{-x}\,\bar{\sigma}_{e\ell}^\uparrow(x)
\end{equation}
which is \emph{exactly} the expected result for a single-fluid plasma.

Using the Klein-Rosseland relation for detailed balance~\cite{oxenius_kinetic_1986},
\begin{equation}\label{eq:mf0_5}
\bar{\sigma}^\uparrow_{s\ell} (x) x \mathpzc{g}_\ell = \bar{\sigma}^\downarrow_{su} (x') x' {\mathpzc{g}}_u
\end{equation}
where $\mathpzc{g}_\ell,\mathpzc{g}_u$ are the degeneracies of the lower and upper atomic levels respectively, we can write the excitation rate as follows:
\begin{equation}\label{eq:mf0_6}
\Gamma^\uparrow_{s\ell} = n_s n_\ell \bar{g}_{\tilde{T}} \; e^{-\lambda} \frac{\mathpzc{g}_u}{\mathpzc{g}_\ell}e^{-x^*} \int_{0}^\infty\!dx'\,x'\,e^{-x'}\,\zeta^{(0)}(\sqrt{\lambda (x^*\!+\!x')})\,\bar{\sigma}_{su}^\downarrow(x')
\end{equation}
One can then easily extract reaction rates, for example:
\begin{subequations}\label{eq:mf0_6b}
\begin{align}
\Gamma^\uparrow_{s\ell} &= \varpi^{\uparrow}_{s\ell}\cdot n_s n_\ell \\
\Gamma^\downarrow_{su} &= \varpi^{\downarrow}_{su}\cdot n_s n_u
\end{align}
\end{subequations}
It is instructive to consider the ratio of these rates:
\begin{equation}\label{eq:mf0_7}
\frac{\varpi^{\uparrow}_{s\ell}}{\varpi^{\downarrow}_{su}} = \left[\frac{\mathpzc{g}_u}{\mathpzc{g}_\ell}e^{-x^*}\right] \cdot 
	\frac{\int_0^\infty\!\!dx'\,x'\,e^{-x'}\zeta^{(0)}(\sqrt{\lambda (x'\!+\!x^*)})\bar{\sigma}^\downarrow_{su}(x')}{\int_0^\infty\!\!dx'\,x'\,e^{-x'}\zeta^{(0)}(\sqrt{\lambda x'})\bar{\sigma}^\downarrow_{su}(x')} 
\end{equation}
The first term in brackets is the traditional Boltzmann equilibrium relation; the second term contains the correction due to the multifluid effects, and appears only through the $\zeta^{(0)}$ function (\ref{eq:ti19}-a). A Taylor expansion near $\lambda\!=\!0$ yields (with an obvious definition of the Boltzmann function $\mathcal{B}$):
\begin{align}\label{eq:mf0_8}
\frac{\varpi^{\uparrow}_{s\ell}}{\varpi^{\downarrow}_{su}} & = \left[\frac{\mathpzc{g}_u}{\mathpzc{g}_\ell}e^{-x^*}\right] \cdot 
	\frac{\int_0^\infty\!\!dx'\,x'\,e^{-x'}\left[1+\frac{2\lambda(x^*\!+\!x')}{3}\right]\bar{\sigma}^\downarrow_{su}(x')}{\int_0^\infty\!\!dx'\,x'\,e^{-x'}\left[1+\frac{2\lambda x'}{3}\right] \bar{\sigma}^\downarrow_{su}(x')} \nonumber \\
	& \simeq  \left[\mathcal{B}_{\ell u}(\tilde{T})\right] \cdot \left(1\!+\!\frac{2\lambda x^*}{3}+o(\lambda^2) \right)
\end{align}
Thus, we recover the expression for Boltzmann equilibrium in the thermal (single-fluid) limit ($\lambda\rightarrow 0$). Note that the correction term increases with the energy threshold, i.e. transitions between high levels ($x^*\rightarrow 0$) will not be affected very much by the multifluid effects, while the impact will be stronger for excitation from low energy levels, with high energy gaps. For elastic collisions ($x^*\!=\! 0$), the ratio of rates is exactly given by the ratio of degeneracies.

\subsection{First-order moment: momentum density}
\label{sec:first-mom}

Consider now the forward reaction (\ref{eq:ti1}) and the corresponding loss of momentum to the particles with velocity $\bvec{v}_s$. The transfer variable in this case is $\psi\!=\! m_s\bvec{v}_s$, and starting from equation (\ref{eq:ti8}), the contribution to the momentum equation is:
\begin{equation}\label{eq:mf1_1}
\begin{split}
\bvec{R}_s^- =  -\frac{n_s n_t}{\pi^{\frac{3}{2}} \alpha^3} \cdot \int d^3 \bvec{V^*} f_{V^*} \cdot  \int dg \, g^3 \, e^{-g^2/\alpha^2} \, \bar{\sigma}_{st}(g) \cdot  \int d\varphi dc_{\theta} \, e^{2gwc_\theta/\alpha^2} \, \avom{m_s \bvec{v}_s}
\end{split}
\end{equation} 
Similarly, the gain in momentum is given by the production of \emph{new} particles with velocity $\bvec{v'}_s$:
\begin{equation}\label{eq:mf1_2}
\begin{split}
\bvec{R}_s^+ = +\frac{n_s n_t}{\pi^{\frac{3}{2}} \alpha^3} \cdot \int d^3 \bvec{V^*} f_{V^*} \cdot  \int dg \, g^3 \, e^{-g^2/\alpha^2} \, \bar{\sigma}_{st}(g) \cdot  \int d\varphi dc_{\theta} \, e^{2gwc_\theta/\alpha^2} \, \avom{m_s \bvec{v'}_s}
\end{split}
\end{equation} 
Using the relation:
\begin{equation}\label{eq:mf1_3}
m_s (\bvec{v}_s - \bvec{v'}_s) = \mu (\bvec{g} - \bvec{g'})
\end{equation} 
we verify that the integrand does not depend on $\bvec{V}^*$ and its integration is trivially removed. The \emph{net} rate of change to the momentum density of species $s$ is therefore:
\begin{equation}\label{eq:mf1_4}
\bvec{R}_s =  -\mu\frac{n_s n_t}{\pi^{\frac{3}{2}} \alpha^3} \cdot  \int dg \, g^3 \, e^{-g^2/\alpha^2} \, \bar{\sigma}_{st}(g) \cdot \int d\varphi dc_{\theta} \, e^{2gwc_\theta/\alpha^2} \, \avom{\bvec{g} \!-\! \bvec{g'}}
\end{equation} 
Let us consider the last integral over the scattering angle. From Figure \ref{fig:coord}, the  vectors $\bvec{g},\bvec{g}'$ in the rotated frame $(\xi,\eta,\varsigma)$ are:
\begin{align}
\bvec{g} = g\ \hat{\bvec{g}} = g\cdot \left(\begin{array}{c} 0\\ 0\\ 1\end{array}\right);
\qquad \bvec{g}' = g'\cdot\ \bvec{\hat{g}}' =g' \left(\begin{array}{c}c_\rho s_\chi\\ s_\rho s_\chi \\ c_\chi\end{array}\right)
\end{align}
Therefore the integral yields:
\begin{equation}\label{eq:mf1_5}
\begin{split}
\int d\Omega' (\bvec{g}\!-\!\bvec{g}') \mathcal{G}(g,\Omega') &= 2\pi g\!\int dc_\chi \mathcal{G}(g,\chi) \hat{\bvec{g}} - 2\pi g'\! \int dc_\chi c_\chi \mathcal{G}(g,\chi) \hat{\bvec{g}} \\
&= \left[ g \!-\! g' \avom{\cos\chi}\right]\,\cart{g}
\end{split}
\end{equation}
We must now express the unit vector $\cart{g}$ in the initial $(\cart{x},\cart{y},\cart{z})$ frame, which is given by (\ref{eq:ti7}); integration over the $\varphi$ variable leaves only one component, $c_\theta \hat{\bvec{w}}$, yielding:
\begin{equation}\label{eq:mf1_6}
\begin{split}
\bvec{R}_s =  -\mu\hat{\bvec{w}}\,\frac{2n_s n_t}{\pi^{\frac{1}{2}} \alpha^3} \cdot  \int dg \, g^3 \, e^{-g^2/\alpha^2} \, \bar{\sigma}_{st}(g) \cdot \int dc_{\theta}c_\theta \, e^{2gwc_\theta/\alpha^2} \, \left[g\!-\!g'\avom{\cos\chi}\right]
\end{split}
\end{equation} 
Using (\ref{eq:ti19}) to replace the last integral and using the normalized variables (\ref{eq:ti12}) leads to:
\begin{align}\label{eq:mf1_7}
\bvec{R}_s = - \frac{2}{3} \mu \bvec{w}\,n_sn_t \bar{g}_{\tilde{T}} e^{-\lambda} \int_{z^*}^\infty dz\,z^{\frac{3}{2}}\,e^{-z}\,\bar{\sigma}_{st}(z)\, \zeta^{(1)} (\sqrt{\lambda z})\left( \sqrt{z} \!-\! \sqrt{z'} \avom{\cos\chi} \right)
\end{align}
A similar (but of opposite sign) expression can be obtained for the species of type $t$, as a result of the identity $m_s (\bvec{v}_t \!-\! \bvec{v'}_t) = \mu (\bvec{g}' \!-\! \bvec{g})$. 

We can now specify the type of collision.
For an $\ell\rightarrow u$ excitation, we follow (\ref{eq:ti15}), to yield:
\begin{align}\label{eq:mf1_8a}
\bvec{R}_s^\uparrow = - \frac{2}{3} \mu \bvec{w}\,n_sn_\ell \bar{g}_{\tilde{T}} e^{-\lambda} \int_{x^*}^\infty dx\,x^{\frac{3}{2}}\,e^{-x}\,\bar{\sigma}_{s\ell}^\uparrow(x)\, \zeta^{(1)} (\sqrt{\lambda x})\left( \sqrt{x} \!-\! \sqrt{x'} \avom{\cos\chi} \right)
\end{align}
For deexcitation, we follow (\ref{eq:ti16}):
\begin{align}\label{eq:mf1_8b}
\bvec{R}_s^\downarrow = - \frac{2}{3} \mu \bvec{w}\,n_sn_u \bar{g}_{\tilde{T}} e^{-\lambda} \int_{0}^\infty dx'\,(x')^{\frac{3}{2}}\,e^{-x'}\,\bar{\sigma}_{su}^\downarrow(x')\, \zeta^{(1)} (\sqrt{\lambda x'})\left( \sqrt{x'} \!-\! \sqrt{x} \avom{\cos\chi} \right)
\end{align}

Note that the expressions (\ref{eq:mf1_8a}--\ref{eq:mf1_8b}) are obtained in a frame where $\bvec{w}$ is aligned with the $\cart{z}$ direction and corresponds to the change in momentum density along that direction. Thus, it is the component of a force \emph{parallel} to $\bvec{w}$, while all components in the transverse directions are zero, by reason of symmetry~\footnote{By integrating over the $\varphi$ angular variable.}. The components in an arbitrary rest-frame must be obtained by projecting $\bvec{w}$.

Since the force density is \emph{approximately} proportional to $\bvec{w}$, we can group all the other terms into the definition of a coefficient, such that
\begin{subequations}\label{eq:mf1_9}
\begin{align}
\bvec{R}^\uparrow_s = -K^\uparrow_{s\ell} (\bvec{u}_s-\bvec{u}_t)  &\qquad \bvec{R}^\uparrow_t = +K^\uparrow_{s\ell} (\bvec{u}_s-\bvec{u}_t)\\
\bvec{R}^\downarrow_s = -K^\downarrow_{su} (\bvec{u}_s-\bvec{u}_t)  &\qquad \bvec{R}^\downarrow_t = +K^\downarrow_{su} (\bvec{u}_s-\bvec{u}_t)
\end{align}
\end{subequations}
where $K^\uparrow_{s\ell}$ and $K^\downarrow_{su}$ are known as resistance coefficients:
\begin{subequations}
\begin{align}\label{eq:mf1_10}
K^\uparrow_{s\ell} &= \frac{2}{3} \mu n_sn_\ell \bar{g}_{\tilde{T}} e^{-\lambda} \int_{x^*}^\infty dx\,x^{\frac{3}{2}}\,e^{-x}\,\bar{\sigma}_{s\ell}^\uparrow(x)\, \zeta^{(1)} (\sqrt{\lambda x})\left[ \sqrt{x} \!-\! \sqrt{x'} \avom{\cos\chi} \right] \\
K_{su}^\downarrow\ &= \frac{2}{3} \mu n_sn_u \bar{g}_{\tilde{T}} e^{-\lambda} \int_{0}^\infty dx'\,(x')^{\frac{3}{2}}\,e^{-x'}\,\bar{\sigma}_{su}^\downarrow(x)\, \zeta^{(1)} (\sqrt{\lambda x'})\left[ \sqrt{x'} \!-\! \sqrt{x} \avom{\cos\chi} \right]
\end{align}
\end{subequations}

It must be pointed out that when the collision is elastic, i.e., $x=x'$, we recover the expression of the momentum transfer cross section often used in transport calculation, i.e., $\sigma^{m} (x) \equiv \bar{\sigma}  (1-\langle \cos \chi \rangle)$ (see for example \cite{hirschfelder_molecular_1964}). In the limit of weak divergence of mean fluid velocities ($\lambda\rightarrow 0$) and isotropic scattering ($\mathcal{G}(\chi)=1/4\pi$), we have:
\begin{align}\label{eq:sc50}
K^\uparrow_{s\ell} \simeq \frac{2}{3}\mu n_s n_\ell \bar{g}_{\tilde{T}}  \int_{x^*}^{\infty} dx x^2\bar{\sigma}_{s\ell}^\uparrow(x)e^{-x}
\end{align}

Again, using the Klein-Rosseland relation, the excitation resistance coefficient can be written as:
\begin{align}\label{eq:mf1_11}
K^\uparrow_{s\ell} = \left[\mathcal{B}_{\ell u}(\tilde{T})\right] \frac{2}{3} \mu\, n_sn_\ell \bar{g}_{\tilde{T}} e^{-\lambda} \int_{0}^\infty dx'\,x'
x^\frac{1}{2}\,e^{-x'}\,\zeta^{(1)} (\sqrt{\lambda (x)}) \bar{\sigma}_{su}^\downarrow(x')\, \left[ \sqrt{x} \!-\!\sqrt{x'} \avom{\cos\chi} \right]
\end{align}
As in the case of the zero-th order moment, we define the momentum exchange rates by:
\begin{subequations}
\begin{align}
K^\uparrow_{s\ell} &= \mu n_s n_\ell \kappa^{\uparrow}_{s\ell}  \\
K^\downarrow_{su} &= \mu n_s n_u \kappa^{\downarrow}_{su} 
\end{align}
\end{subequations}
In the case of weak divergence of mean fluid velocities and isotropic scattering, the ratio of the rate coefficients for the forward and backward processes is approximately:
\begin{equation}\label{eq:mf1_12}
\frac{\kappa^{\uparrow}_{s\ell} }{\kappa^{\downarrow}_{su}} \simeq \left[ \mathcal{B}_{\ell u} (\tilde{T})\right] \cdot \frac{\int_{0}^{\infty} dx'e^{-x'} x'(x^*\!+\!x') \left[ 1+\frac{2}{5} \lambda (x^*\!+\!x')\right] \bar{\sigma}_{su}^\downarrow(x')}{\int_{0}^{\infty} dx'e^{-x'} x'^2 \left[ 1+\frac{2}{5} \lambda x'\right] \bar{\sigma}_{su}^\downarrow(x')} 
\end{equation}
Note that there is an additional contribution from high-order moment from the expansion. This can be seen by further expanding the integrand of the numerator:
\begin{equation*}
x'(x^*\!+\!x') \left[ 1+\frac{2}{5} \lambda (x^*\!+\!x')\right] = x'^2 \left[ 1+\frac{2}{5} \lambda x'\right] + x^*x'  \left[ 1+\frac{2}{5} \lambda (x^*+2x')\right]
\end{equation*}
such that
\begin{equation}\label{eq:mf1_13}
\frac{\kappa^{\uparrow}_{s\ell} }{\kappa^{\downarrow}_{su}} \simeq \left[ \mathcal{B}_{\ell u} (\tilde{T})\right] \cdot 
\left[ 1 + x^* 
\frac{\int_{0}^{\infty} dx'e^{-x'} x'\left[ 1+\frac{2}{5} \lambda (x^*\!+\!2x')\right] \bar{\sigma}_{su}^\downarrow(x')}{\int_{0}^{\infty} dx'e^{-x'} x'^2 \left[ 1+\frac{2}{5} \lambda x'\right] \bar{\sigma}_{su}^\downarrow(x')} 
\right]
\end{equation}
Note that even as $\lambda\rightarrow 0$, the correction term does not vanish.
Therefore, there is no equivalence between the resistance coefficients of the forward and backward processes in the limit $\lambda \rightarrow 0$. However, this is perfectly understandable; note that the correction is proportional to the energy threshold, and since kinetic energy must be removed from particle $s$ in order to achieve excitation, but not for deexcitation, there must also be an imbalance in the momentum exchange rate. As expected, this imbalance vanishes for elastic collisions ($x^*\rightarrow 0$), and the rates are consistent with the detailed balance of the mass exchange. In all cases, detailed balance is enforced through relation (\ref{eq:mf0_5}) at the microscopic level. 

\subsection{Second-order moment: total energy density}
\label{sec:second-mom}

The net rate of change of total energy of species $s$ can be obtained by setting $\psi \!=\! \frac{1}{2} m_s \left( \bvec{v'}^2_s \!-\! \bvec{v}^2_s \right)$ into equation (\ref{eq:ti11}):
\begin{equation}\label{eq:mf2_1}
	Q_s= \frac{n_s n_t}{\pi^{\frac{3}{2}} \alpha^3} \cdot \int d^3 \bvec{V^*} f_{V^*} \int dg \, g^3 \, e^{-g^2/\alpha^2} \, \bar{\sigma}_{st}(g) \cdot  \int d\varphi dc_\theta \, e^{2gwc_\theta/\alpha^2} \, \avom{\frac{1}{2} m_s (\bvec{v'}_s^2 \!-\! \bvec{v}_s^2) }
\end{equation}
Using the transformation defined in Appendix~\ref{app:B}, 
\begin{align}\label{eq:mf2_2}
\frac{1}{2} m_s \left( \bvec{v'}_s^2 \!-\! \bvec{v}_s^2 \right) = \mu \left( \bvec{g'} \!-\! \bvec{g} \right) \cdot \left[ \bvec{V^*} \!+\! \bvec{U} \!-\! \gamma (\bvec{g} \!-\! \bvec{w}) \right] - \frac{m_t}{M} \Delta\varepsilon
\end{align}
The integration of the first term in the square bracket is zero since $\int d^3\bvec{V^*} \, \bvec{V^*} \, f_{V^*} = 0$. 
One can easily see that the second term in brackets is simply $\bvec{R}_s \cdot \bvec{U}$ by comparing with (\ref{eq:mf1_4}). Similarly, the last term in (\ref{eq:mf2_2}) is identified as $-(m_t/M)\Delta\varepsilon \Gamma_{st}$. The third term in brackets involves the following dot product:
\begin{align}\label{eq:mf2_3}
(\bvec{g}'\!-\!\bvec{g})\cdot(\bvec{g}\!-\!\bvec{w}) &= \bvec{g}\cdot\bvec{g}' -\bvec{w}\cdot\bvec{g}' - g^2 +\bvec{w}\cdot\bvec{g}\nonumber\\
                                        &= gg'\cos\chi -wg'(\hat{\bvec{g}}'\cdot\hat{\bvec{w}}) -g^2 +wg\cos\theta
\end{align}
We can now perform the averaging over the scattering angle; in particular, we have
\begin{equation*}
\int\! d\Omega' \mathcal{G}(\chi)\,\hat{\bvec{g}}'\!\cdot\!\hat{\bvec{w}} = \avom{\cos\chi}\cos\theta
\end{equation*}
so that
\begin{align}\label{eq:mf2_4}
\int\! d\Omega' \mathcal{G}(\chi)\,(\bvec{g}'\!-\!\bvec{g})\!\cdot\!(\bvec{g}\!-\!\bvec{w}) = (g'\avom{\cos\chi}\!-\!g)(g\!-\!w\cos\theta)
\end{align}
Thus, we obtain, after integration over $\varphi$:
\begin{align}\label{eq:mf2_5}
Q_s  = \bvec{U}\cdot\bvec{R}_s -\frac{m_t}{M}\Delta\varepsilon\Gamma_{st} 
- \mu\gamma\frac{4n_sn_t}{\pi^\frac{1}{2}\alpha^3} e^{-\lambda}\int\!\!dg\,g^3\bar{\sigma}_{st} (g) e^{-g^2/\alpha^2}\,\mathcal{I}_\theta
\end{align}
where the last angular integral is
\begin{align}\label{eq:mf2_6}
\mathcal{I}_\theta &= \frac{1}{2}\int\!dc_\theta e^{2gwc_\theta/\alpha^2} (g'\avom{c_\chi}\!-\!g)(g\!-\!wc_\theta)\nonumber\\
             &= (g'\avom{c_\chi}\!-\!g) \left\{ g\frac{1}{2}\int\!dc_\theta e^{2gwc_\theta/\alpha^2} -\frac{w}{2}\int\!dc_\theta c_\theta e^{2gwc_\theta/\alpha^2} \right\}\\
             &= (g'\avom{c_\chi}\!-\!g)\,g\, \left\{\zeta^{(0)}(\sqrt{\lambda z}) 
             -\frac{2}{3}\frac{w^2}{\alpha^2} \zeta^{(1)}(\sqrt{\lambda z}) \right\}\nonumber
\end{align}

Therefore, after the change of variables $g\rightarrow z$:
\begin{align}\label{eq:mf2_7}
Q_s  = & \bvec{U}\cdot\bvec{R}_s   - \frac{m_t}{M}\Delta\varepsilon\Gamma_{st} \nonumber\\
 & + \mu\gamma\frac{4n_sn_t}{\pi^\frac{1}{2}\alpha^3} e^{-\lambda}\frac{\alpha^6}{2} 
  \int\!\!dzz^\frac{3}{2} e^{-z} \bar{\sigma}_{st} (z)
 \left(\sqrt{z}\!-\!\sqrt{z'}\avom{\cos\chi}\right) 
 \left(\zeta^{(0)}(\sqrt{\lambda z}) \!-\! \frac{2\lambda}{3}\zeta^{(1)}(\sqrt{\lambda z}) \right)
\end{align}
The factor in front of the integral can be re-arranged to yield:
$$\gamma n_s n_t \bar{g}_{\tilde{T}} e^{-\lambda} (2k\tilde{T})$$
Using the identity $
\gamma(2k\tilde{T}) = \frac{2\mu}{M} k(T_t-T_s)
$, the final result has a traditional form:
\begin{align}\label{eq:mf2_8}
Q_s  = \bvec{U}\cdot\bvec{R}_s -\frac{m_t}{M}\Delta\varepsilon\Gamma_{st} +J_{st}\frac{2\mu}{M} k(T_t-T_s)
\end{align}
with the thermal resistance coefficient defined as:
\begin{align}\label{eq:mf2_9}
J_{st} = n_s n_t \bar{g}_{\tilde{T}} e^{-\lambda} \int\!\!dzz^\frac{3}{2} e^{-z} \bar{\sigma}_{st} (z)
\left(\sqrt{z}\!-\!\sqrt{z'}\avom{\cos\chi}\right) \left(\zeta^{(0)}(\sqrt{\lambda z}) - \frac{2\lambda}{3}\zeta^{(1)} (\sqrt{\lambda z}) \right)
\end{align}

This result is general, and we can now make the usual substitutions for excitation:
\begin{subequations}\label{eq:mf2_10}
\begin{align}
Q_s^\uparrow  &= \bvec{U}\cdot\bvec{R}^\uparrow_s -\frac{m_t}{M}\varepsilon^*\Gamma^\uparrow_{s\ell} +J^\uparrow_{s\ell} \frac{2\mu}{M} k(T_t-T_s)\\
J^\uparrow_{s \ell} &= n_s n_\ell \bar{g}_{\tilde{T}} e^{-\lambda} \int\!\!dxx^\frac{3}{2} e^{-x} \bar{\sigma}^\uparrow_{s\ell} (x) 
\left(\sqrt{x}\!-\!\sqrt{x'}\avom{\cos\chi}\right) \left(\zeta^{(0)}(\sqrt{\lambda x}) - \frac{2\lambda}{3}\zeta^{(1)}(\sqrt{\lambda x})\right)
\end{align}
\end{subequations}
In the case of deexcitation, we can still use the general formula (\ref{eq:mf2_7}), except that in this case, $\Delta\varepsilon=-\varepsilon^*$. This can be seen if we start from eq. (\ref{eq:mf2_1}), which gives us:
\begin{align}\label{eq:mf2_11}
Q_s^\downarrow = &-\mu\frac{n_sn_t}{\pi^\frac{3}{2}\alpha^3} e^{-\lambda} \int d^3\bvec{g}' g' \bar{\sigma}^\downarrow_{su} (g') e^{-g'^2/\alpha^2} e^{2g'wc_\theta/\alpha^2}\bvec{U}\cdot\langle \bvec{g}'\!-\!\bvec{g}\rangle_{_{\Omega'}}\nonumber\\
 & +\mu\gamma \frac{n_sn_t}{\pi^\frac{3}{2}\alpha^3} e^{-\lambda}\int d^3\bvec{g}' g' \bar{\sigma}^\downarrow_{su} (g') e^{-g'^2/\alpha^2} e^{2g'wc_\theta/\alpha^2} (\bvec{g}'\!-\!\bvec{g})\cdot(\bvec{g}\!-\!\bvec{w})\\
 & +\frac{m_t}{M}\varepsilon^* \Gamma_{su}^\downarrow\nonumber
\end{align}
Again, one can easily recognize the standard formulae:
\begin{subequations}\label{eq:mf2_12}
\begin{align}
Q_s^\downarrow  &= \bvec{U}\cdot\bvec{R}^\downarrow_s +\frac{m_t}{M}\varepsilon^*\Gamma^\downarrow_{su} + J^\downarrow_{su}\frac{2\mu}{M} k(T_t-T_s)\\
J^\downarrow_{su} &= n_s n_u \bar{g}_{\tilde{T}} e^{-\lambda} \int\!\!dx'(x')^\frac{3}{2} e^{-x'} \bar{\sigma}^\downarrow_{su} (x')
\left(\sqrt{x'}\!-\!\sqrt{x}\avom{\cos\chi}\right) \left(\zeta^{(0)}(\sqrt{\lambda x'}) - \frac{2\lambda}{3}\zeta^{(1)}(\sqrt{\lambda x'})\right)
\end{align}
\end{subequations}
which could also be obtained directly from (\ref{eq:mf2_8} - \ref{eq:mf2_9}), with the usual substitutions (\ref{eq:ti16}).

We can also express the source term for particle $t$, using:
\begin{align}\label{eq:mf2_13}
\frac{1}{2}m_t(\bvec{v}_t'^2\!-\!\bvec{v}_t^2) &= \frac{1}{2}m_t\left( (\bvec{V}\!-\!\frac{m_s}{M}\bvec{g}')^2 - (\bvec{V}\!-\!\frac{m_s}{M}\bvec{g})^2\right)\nonumber\\
 &= \frac{m_s}{2M}\mu (\bvec{g}'^2\!-\!\bvec{g}^2) -\mu \bvec{V}\cdot(\bvec{g}'\!-\!\bvec{g})\\
 &= -\mu(\bvec{g}'\!-\!\bvec{g}) \cdot \left[\bvec{V}^*+\bvec{U}-\gamma(\bvec{g}\!-\!\bvec{w}) \right] -\frac{m_s}{M}\Delta\varepsilon \nonumber
 \end{align}
 Comparing with (\ref{eq:mf2_2}), we easily obtain (note the inversion of $T_s$ and $T_t$ in the last term):
\begin{subequations}\label{eq:mf2_14}
 \begin{align}
Q_t^\uparrow  &= \bvec{U}\cdot\bvec{R}^\uparrow_t -\frac{m_s}{M}\varepsilon^*\Gamma^\uparrow_{s\ell} +J^\uparrow_{s\ell}\,\frac{2\mu}{M} k(T_s-T_t) \\
Q_t^\downarrow  &= \bvec{U}\cdot\bvec{R}^\downarrow_t +\frac{m_s}{M}\varepsilon^*\Gamma^\downarrow_{su} +J^\downarrow_{su}\,\frac{2\mu}{M} k(T_s-T_t)
\end{align}
\end{subequations}
with $\bvec{R}_t^{\uparrow (\downarrow)} = -\bvec{R}_s^{\uparrow (\downarrow)}$.
Combining both $s$ and $t$ fluids, the only term remaining is the loss of energy equal to the energy gap between the levels, as expected. Note also that this energy loss is distributed to the respective fluids according to the ratio of masses, such that the lighter element receives the major contribution. This is also an expected result, similar to the energy exchange due to elastic collisions, and due to the kinematics of collision.

In the limits of near single-fluid ($\lambda\rightarrow 0$) and isotropic scattering, we have:
\begin{align}\label{eq:mf2_15}
J^\uparrow_{s \ell} &\simeq n_s n_\ell \bar{g}_{\tilde{T}} \left(1 - \frac{2\lambda}{3}\right) \int_{x^*}^{\infty} dx\,x^2\, \bar{\sigma}^\uparrow_{s\ell} (x) e^{-x}
\end{align}
The thermal relaxation rates can be extracted similarly:
\begin{subequations}
\begin{align}
J^\uparrow_{s \ell} &= n_s n_\ell j^\uparrow_{s \ell} \\
J^\downarrow_{s u} &= n_s n_u j^\downarrow_{s u}
\end{align}
\end{subequations}
The ratio of the thermal relaxation rates can be written as:
\begin{align}
\frac{j^\uparrow_{s \ell}}{j^\downarrow_{s u}} \simeq \left[ \mathcal{B}_{\ell u} (\tilde{T}) \right] \cdot \left[ 1+ x^* \frac{\int_0^\infty dx' \, e^{-x'} \, x' \, \left[ 1 - \frac{2}{3} \lambda + \frac{2}{3} \lambda (x^*+2x') \right] \, \bar{\sigma}^\downarrow_{su} }{\int_0^\infty dx' \, e^{-x'} \, x'^2 \,\left[ 1 - \frac{2}{3} \lambda + \frac{2}{3} \lambda x' \right] \, \bar{\sigma}^\downarrow_{su}} \right]
\end{align}
Similar to the case of momentum transfer rates, the correction terms do not vanish when $\lambda \rightarrow 0$ due to contribution from high-order moments.

\section{Numerical Results}
\label{sec:numerix}
In the following sections, we carry out a numerical evaluation and verification of the exchange rates derived in \ref{sec:zero-mom}, \ref{sec:first-mom} and \ref{sec:second-mom}, for the case of free electrons interacting with hydrogen atoms; these processes include electron-neutral elastic collision and electron-impact excitation and deexcitation. Ionization and recombination are currently omitted and will be included in future work. For comparison purpose, we also show the exchange rates due to Coulomb collision, i.e., electron-hydrogen ion, which is the dominating elastic exchange mechanism for plasma with high ionization fraction. 

The notations are slightly modified to better distinguish each type of interaction. We use superscripts $(en)$, $(ei)$ and $(xd)$ to denote electron-neutral, electron-ion (Coulomb), and excitation/deexcitation (as a whole) collisions, respectively. These grouped notations are useful, for example, when looking at the net momentum (or energy) transfer due to each type of collision. The symbols $\uparrow$, $\downarrow$ are still retained to indicate individual excitation and deexcitation rates. For each transition between two atomic states, we use the convention of indexing the final state on the left, and the initial state on the right, i.e., $(f|i)$. For example, $\varpi^\uparrow_{(u|\ell)}$ is the forward excitation rate from $\ell$ to $u$, and $\varpi^\downarrow_{(\ell|u)}$ is the reverse process.
The energy levels and cross sections models for atomic hydrogen are given in classical form and summarized in Appendix \ref{app:F}.

\subsection{Reaction Rates}
\label{sec:numerix-1}
All the exchange rates (mass, momentum, energy) can be tabulated as a function of two parameters: the average thermal temperature $\tilde{T}$, defined in appendix \ref{app:B}, and a non-dimensional parameter $\lambda$, defined in eq.~(\ref{eq:ti12}), which corresponds to the relative mean kinetic energy. For convenience, we also define an equivalent drift temperature $T_w = \lambda \tilde{T}$\footnote{Equivalently, one can also define $kT_w = \frac{1}{2}\mu w^2$}, such that all the rates can be tabulated in terms of two temperatures. Since the mass ratio between the electron and the atom is very small ($m_e \ll M_H$), we can drop terms of order $m_e/M$ and arrive at the following approximations: $\mu \simeq m_e$, $\tilde{T} \simeq T_e$, $T_w \simeq \lambda T_e$ and $\bar{g}_{\tilde{T}} \simeq \bar{v}_e = \sqrt{ \frac{8 kT_e}{\pi m_e}}$. Therefore, all the exchange rates for electron-induced collisions (both elastic and inelastic) can be numerically evaluated in terms of the electron temperature $T_e$ and the drift temperature $T_w$.  

\begin{figure}
\centering
\includegraphics[scale=1]{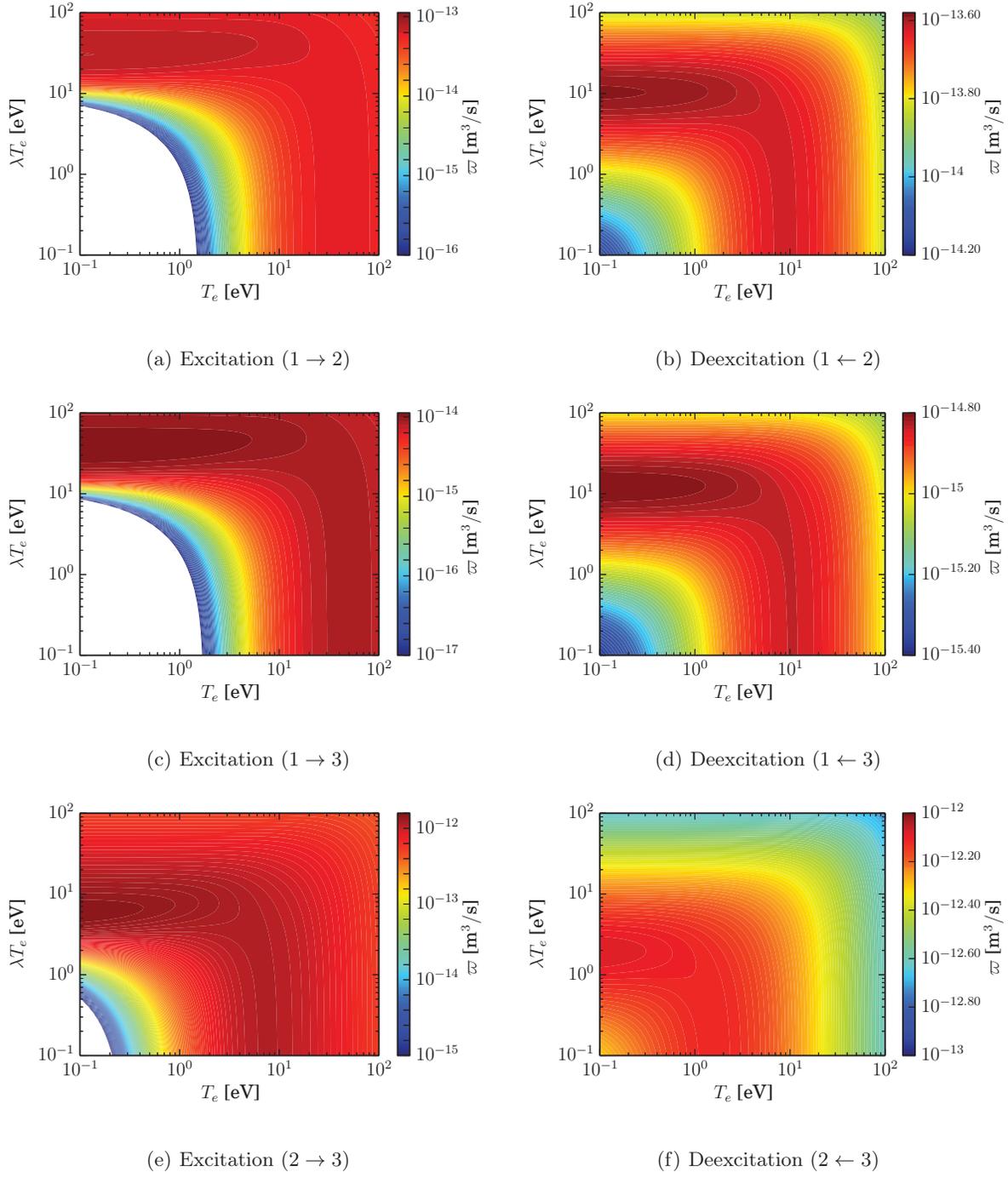}
\caption{Multifluid reaction rates for electron-impact excitation/deexcitation collisions as a function of two temperatures. For better color contrast, all the values smaller than minimum value of the colormap are white out.}
\label{fig:ED1}
\end{figure}

Figure~\ref{fig:ED1} shows example calculations of the zeroth-order reaction rates, defined in eqs.~(\ref{eq:mf0_2a}), (\ref{eq:mf0_2b}) and (\ref{eq:mf0_6b}), for electron-impact excitation and deexcitation between the first three atomic states of hydrogen. These rates exhibit a similar trend for the range of temperatures plotted here, that is, starting from low temperature, the rates first increase, reaching a plateau and then decrease as temperature further increases. The value at which the rate is maximum is very close to the threshold temperature of the transition. This trend holds both in the direction of increasing thermal $T_e$ (x-axis) or drift temperatures $T_w$ (y-axis).

It is clearly shown from Figure~\ref{fig:ED1} that the reaction rates can be significantly different from the thermal limit when the relative mean velocity between two fluids is significant. In particular, one sees an increase of the reaction rate in the low temperature regime where $T_e$ and $T_w$ are small compared to the excitation temperature of the collision; this enhancement corresponds directly to the form of the cross sections. Therefore, one can expect that significant deviation from the thermal rate occurs when the mean kinetic energy is of the same order as the excitation temperature. This indicates that excitation and deexcitation among low energy states with large threshold energies are more sensitive to the multifluid effects. This is consistent with the prior statement we made when examining the ratio of forward and backward rates.

\begin{figure}
\centering
\includegraphics[height=7cm]{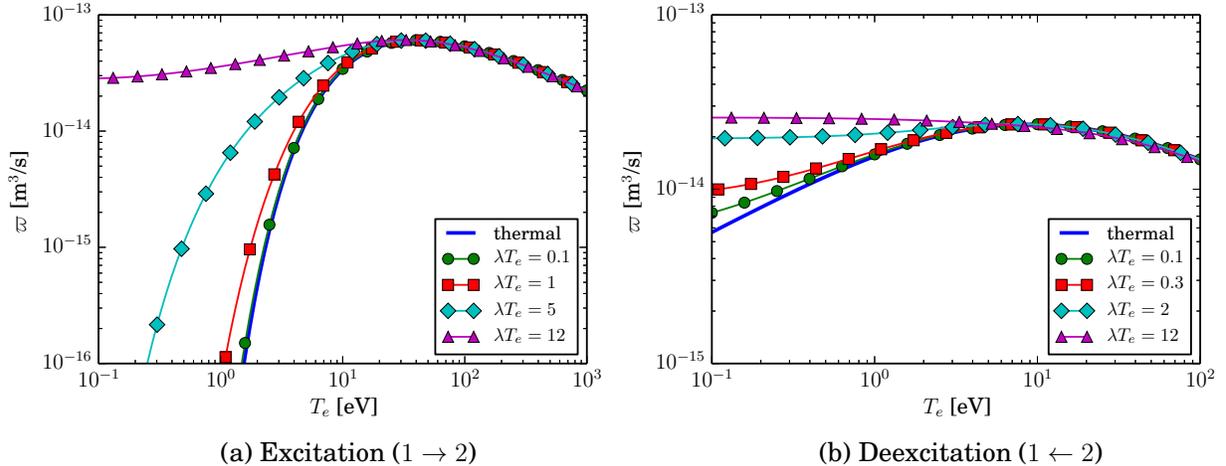}
\caption{Multifluid reaction rates for electron-impact excitation/deexcitation collisions: lines with symbols correspond to different values of the drift temperature. The solid line corresponds to the exact solution for the thermal limit given by equation~(\ref{eq:AF-1}).}
\label{fig:ED2}
\end{figure}

\begin{figure}
\centering
\includegraphics[height=7cm]{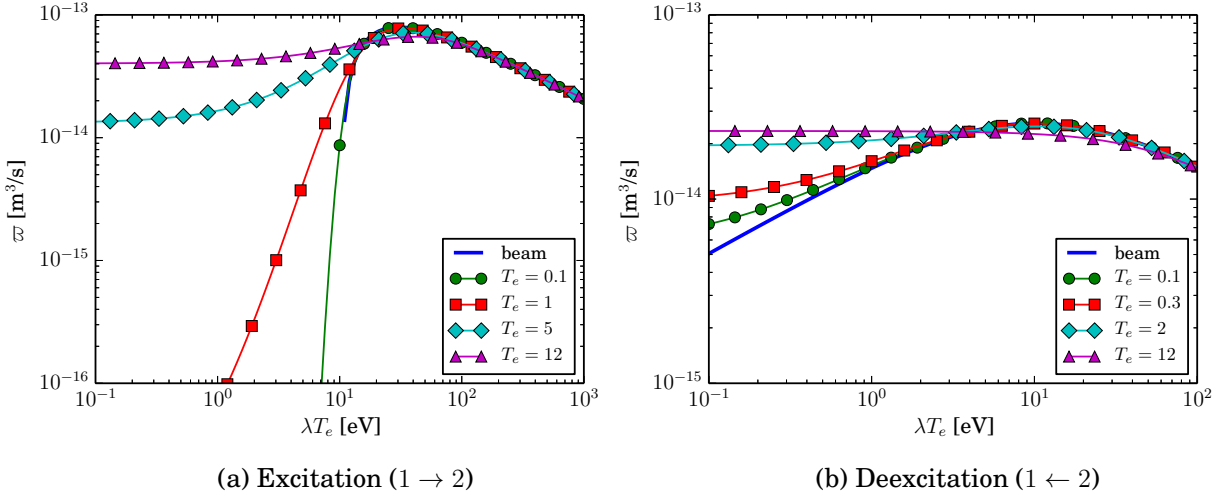}
\caption{Multifluid reaction rates for electron-impact excitation/deexcitation collisions: lines with symbols correspond to different values of the thermal temperature. The solid line corresponds to the exact solution for the beam limit given by equation~(\ref{eq:AF-2}).}
\label{fig:ED3}
\end{figure}

Figure~\ref{fig:ED2} shows the forward and backward reaction rates of the first transition between the ground state and the first excited state as a function of the \emph{thermal} temperature for several drift temperature values. Even at very low thermal temperature ($T_e \simeq 0.1$ eV), one can have significant excitation ($\varpi \simeq 10^{-14}$ m$^3/$s) due to a high drift temperature. In addition, Figure~\ref{fig:ED2} also shows that in the limit $\lambda \rightarrow 0$, the multifluid rate, as formulated here, converges to the expression for thermal limit, given in eq.~(\ref{eq:AF-1}), as expected. Figure~\ref{fig:ED3} shows the reaction rates as a function of the \emph{drift} temperature for several thermal temperature values; here we can identify a different asymptotic limit of the rate. In the limit $\lambda \rightarrow \infty$, the electron velocity distribution function approaches the form of a delta function centered at the relative mean velocity $\mathbf{w}$, and the reaction rates approach the beam limit given by eq.~(\ref{eq:AF-2}). It must be noted that in the numerical integration of the multifluid rates, e.g., eq.~(\ref{eq:mf0_2a}), the energy grid $x$ needs to be refined near the value of the the mean kinetic energy $\lambda$ to avoid numerical error due to the integration over a delta function.

\subsection{Momentum and Energy Exchange Rates}
\label{sec:numerix-2}
We now compute the momentum and energy exchange rates due to both excitation and deexcitation, and compare with those due to elastic collisions (electron-neutral and electron-ion). Recall from eq.~(\ref{eq:mf2_8}) that the total energy transfer include three terms: the first term due to work done by friction in the COM reference frame, the second term due to thermal resistance (or thermal relaxation), and the last one due to heat release/absorption due to chemical reaction. There are three different rate coefficients associated with each of these processes, namely the momentum exchange rate $\kappa$, thermal relaxation rate $j$, and the reaction rate $\varpi$. In this section, we will focus on examining the momentum exchange and thermal relaxation rates. The expressions derived in \ref{sec:first-mom} and \ref{sec:second-mom} can be readily used for the case of elastic collisions by simply setting $z^* = 0$ and $z'=z=x$. For example, in the case of electron-neutral $(en)$ collision, we have:
\begin{subequations}
\begin{align}
\kappa^{(en)} &= \frac{2}{3} \bar{v}_e e^{-\lambda} \int_{0}^\infty dx\,x^2\,e^{-x}\,\bar{\sigma}^{(en)}(x)\, \left[ 1 \!-\! \avom{\cos\chi} \right] \zeta^{(1)} (\sqrt{\lambda x})\\
j^{(en)} &= \bar{v}_e e^{-\lambda} \int_0^\infty dx\, x^2 e^{-x} \bar{\sigma}^{(en)}(x) \left[ 1 \!-\! \avom{\cos\chi} \right]
 \left[\zeta^{(0)}(\sqrt{\lambda x}) - \frac{2\lambda}{3}\zeta^{(1)} (\sqrt{\lambda x})  \right]
\end{align} 
\label{eq:n-1}
\end{subequations}

It can be seen from the previous two equations that in order to compute the rate for the case of elastic collisions, we only need the so-called momentum transfer cross section $\sigma^{(en)m} (x) \equiv \bar{\sigma}^{(en)}  (1-\langle \cos \chi \rangle)$. These cross sections are available for a wide range of neutral species due to their extensive use in calculation of transport properties (see for example \cite{capitelli_fundamental_2013}). The electron-neutral collision cross section utilized in this work is taken from Bray and Stelbovics \cite{bray_convergent_1992}. For inelastic collisions, we need the full DCS, i.e., both $\bar{\sigma} (x)$ and $\mathcal{G} (x,\chi)$ in eq.~(\ref{eq:ti9}) for each process. These cross sections are generally not available and analytical approximation is needed.
For Coulomb collision, we use the analytical DCS from Rutherford's scattering formula with suitable cut-off based on the Debye length \cite{biberman_kinetics_1987}, yielding a momentum transfer cross section of the form:
\begin{align}
\sigma^{(ei)m} = \frac{e^4}{ 16 \pi \epsilon_0^2 \varepsilon^2} \ln \Lambda
\end{align}
where $\ln \Lambda$ is the well-known Coulomb logarithm,
\begin{align*}
\Lambda = 1.24 \times 10^7  \left( \frac{T_e^3}{n_e} \right)^{1/2}
\end{align*}
Typically, $\ln \Lambda \approx 5-20$; for convenience, we take a constant value of $\ln \Lambda = 5$ for all the plots shown here. In this case, the momentum exchange and thermal relaxation rates can be obtained in exact forms \cite{burgers_flow_1969}:
\begin{subequations}\label{eq:kCoul}
\begin{align}
\kappa^{(ei)} (T_e,\lambda) &= \frac{e^4 \ln \Lambda}{16\pi \epsilon_0^2 (kT_e)^2}  \frac{1}{\lambda} \left[ \frac{\textrm{erf} (\sqrt{\lambda}) }{\sqrt{\lambda}} - \frac{2}{\sqrt{\pi}} e^{-\lambda} \right]\\
j^{(ei)} (T_e,\lambda) & = \frac{e^4 \ln \Lambda}{16\pi \epsilon_0^2 (kT_e)^2} \left[   \frac{2}{\sqrt{\pi}} e^{-\lambda} \right]
\end{align}
\end{subequations}
where $\textrm{erf}$ is the typical error function.

\begin{figure}
\centering
\includegraphics[scale=.85]{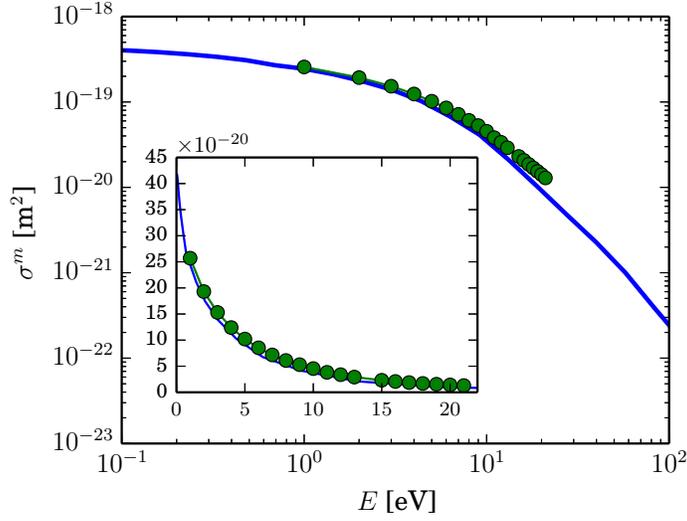}
\caption{Elastic momentum transfer cross section for electrons in atomic hydrogen computed with the DCS from equation~(\ref{eq:DCS}). The symbols are theoretical values from Bray and Stelbovics \cite{bray_convergent_1992}.}
\label{fig:xsms}
\end{figure}

Due to the lack of data of the DCS for inelastic processes, we have used an analytical Born scattering approximation for a Coulomb screened potential \cite{capitelli_fundamental_2013}, given by the following form:
\begin{align}
\label{eq:DCS}
\mathcal{G} (\varepsilon,\chi) = \frac{C}{(1-h\cos \chi)^2}; \quad h = \frac{1}{1+\frac{21.8}{\varepsilon}}
\end{align}
where $C$ is a normalization constant such that $2\pi \int_{-1}^1 \mathcal{G} (\varepsilon,\chi) d c_\chi = 1$. This angular-dependent DCS has been used to compute momentum transfer and thermal relaxation rates for both electron-neutral and excitation/deexcitation collisions. We note here that the angular-dependent DCS's $\mathcal{G}$ for excitation and deexcitation have to satisfy detailed balance \cite{oxenius_kinetic_1986}:
\begin{align}
\mathcal{G}^\uparrow_{(u|\ell)} (\varepsilon,\chi) = \mathcal{G}^\downarrow_{(\ell|u)} (\varepsilon',\chi)
\end{align}
The above condition implies that the probability for a deexcitation collision with an incident energy $\varepsilon$ to have a scattering angle of $\chi$ is the same as that for an excitation collision with an incident energy $\varepsilon+\varepsilon^*$; therefore, one cannot independently specify both DCS's for the forward and backward processes. Figure \ref{fig:xsms} shows a comparison of the computed elastic momentum transfer cross sections to the result from a direct close-coupling calculation of Bray and Stelbovics \cite{bray_convergent_1992}; the agreement between the two is excellent. 
Using the angular-dependent DCS defined in eq.~(\ref{eq:DCS}), the exchange rates, e.g., $\kappa$ and $j$, can be computed for each bound-bound transition and summed over all transitions to yield the total rates:
\begin{subequations}
\begin{align}
K^{(xd)} &= m_e \sum_\ell \sum_{u > \ell}  \left( \kappa^\uparrow_{(u|\ell)} n_l n_e + \kappa^\downarrow_{(\ell|u)} n_u n_e \right)\\
J^{(xd)} &=  \sum_\ell \sum_{u > \ell} \left(  j^\uparrow_{(u|\ell)} n_l n_e + j^\downarrow_{(\ell|u)} n_u n_e \right)
\end{align}
\label{eq:n-2}
\end{subequations}
Based on the total frictional and thermal resistance coefficients, we can extract \emph{average} momentum transfer and thermal relaxation rates for all excitation/deexcitation processes as follows:
\begin{subequations}
\begin{align}
K^{(xd)} &= m_e n_e n_n \kappa^{(xd)}\\
J^{(xd)} &= n_e n_n j^{(xd)}
\end{align}
\label{eq:n-3}
\end{subequations}
where $n_n = \sum_{k} n_k$ is the total atomic number density (summation over levels). Note that according to our definitions, the exchange rates $\kappa^{(xd)}$ and $j^{(xd)}$ contain terms designating the population of the excited states, e.g., $n_k/n_n$. For comparison purpose, these average rates are calculated by assuming a Boltzmann distribution of the atomic states, i.e., $n_k= n_n \frac{g_k e^{-E_k/kT_B}}{\Zcal_n}$ and $\Zcal_n = \sum_{k \in n} g_k e^{-E_k/kT_B}$. One can see that in this case the population of the excited states is effectively replaced by a Boltzmann distribution characterized by a temperature $T_B$. This step is only done for the comparison shown below. In a CR calculation, the \emph{detailed} population of the atomic states (equilibrium or not) is known and the exchange rates are computed for each transition as specified in eq. (\ref{eq:n-2}).

\begin{figure}
\centering
\includegraphics[scale=1]{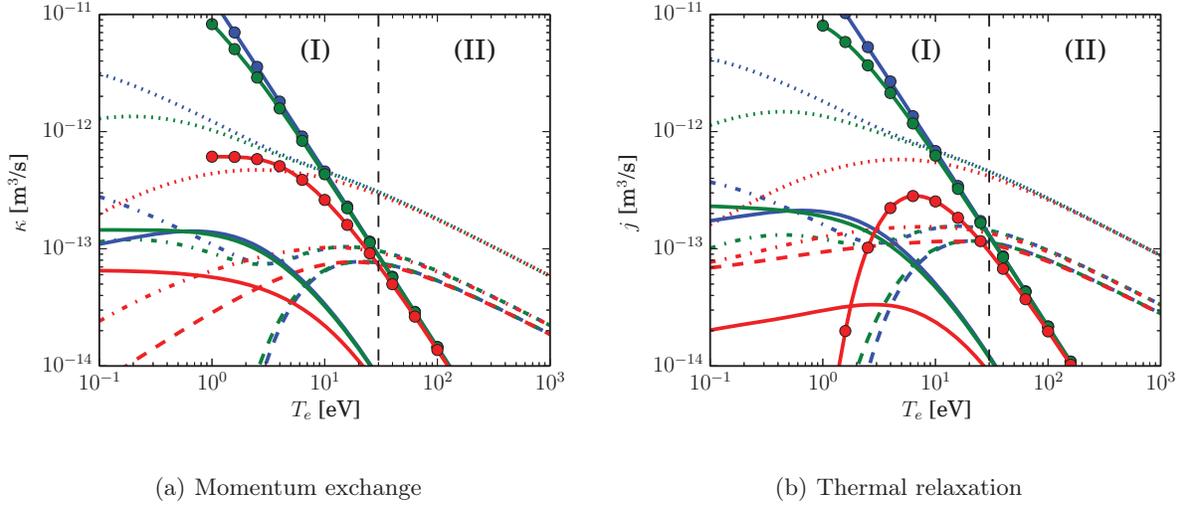}
\caption{Comparison of the momentum exchange rates $\kappa$ and thermal relaxation rates $j$ due to elastic and inelastic collisions (excitation and deexcitation). Different colors of the line indicate different values of the drift temperatures in eV: $T_w = 0.1$ (blue), $1$ (green) and $10$ (red) eV. Solid lines indicate rates due to e$^-$-H elastic collisions, solid lines with circles indicate e$^-$-H$^+$ (Coulomb) collisions, and three sets of broken lines are used for the inelastic collisions. The rates for inelastic collisions are defined in eq.~(\ref{eq:n-3}) and are: dashed, dash-dotted and dotted lines for the case of $T_B = 0.5,0.7$ and $0.8$ eV, respectively.}
\label{fig:elas_inel}
\end{figure}

Figure \ref{fig:elas_inel} shows a comparison between the momentum exchange and thermal relaxation rates for three different processes: electron-neutral, electron-ion (Coulomb), and excitation/deexcitation collisions. For clarity, the rates due to Coulomb collision are only shown for value of $T_e > 1$ eV. Two important observations can be deduced from this plot. Firstly, when the atoms are cold, i.e., $T_B$ is low, the inelastic exchange rates are much smaller than elastic ones. This is due to the fact that when $T_B$ is low, only collisions between the ground state and a first few excited states are significant; the rates for these transitions are low compared to the others due to larger energy threshold. As the atoms are being excited and $T_B$ increases, transitions among highly excited states become significant, leading to an overall increase in the total rates. These rates eventually exceed those due to elastic collisions as can be noticed in region (I) in Figure \ref{fig:elas_inel} for the dash-dotted and dotted lines. Secondly, the rates due to inelastic processes tends to have a slower drop-off at high temperature compared to the elastic rates, which suggests that at sufficiently high temperature, the main momentum and energy transfer mechanisms (region (II) in Figure \ref{fig:elas_inel}) are due to inelastic collisions. In addition, one can also observe that the thermal relaxation rates decrease at low thermal and high drift temperatures (red solid curve with and without circles). This is due the cancellation of two terms inside the last bracket of eq.~(\ref{eq:mf2_9}); the second term has a multiplicative factor of $\lambda$ hence its magnitude is increased at high drift temperature. Thus, we are led to the important conclusion that one \emph{cannot} neglect momentum and energy transfer due to inelastic collisions. The only justification for neglecting these terms is when the atoms are cold and thermal temperature is low, but both of these conditions will not be realized in most practical systems.

\subsection{Verification}
\label{sec:numerix-3}

\begin{figure}
\centering
\includegraphics[scale=1]{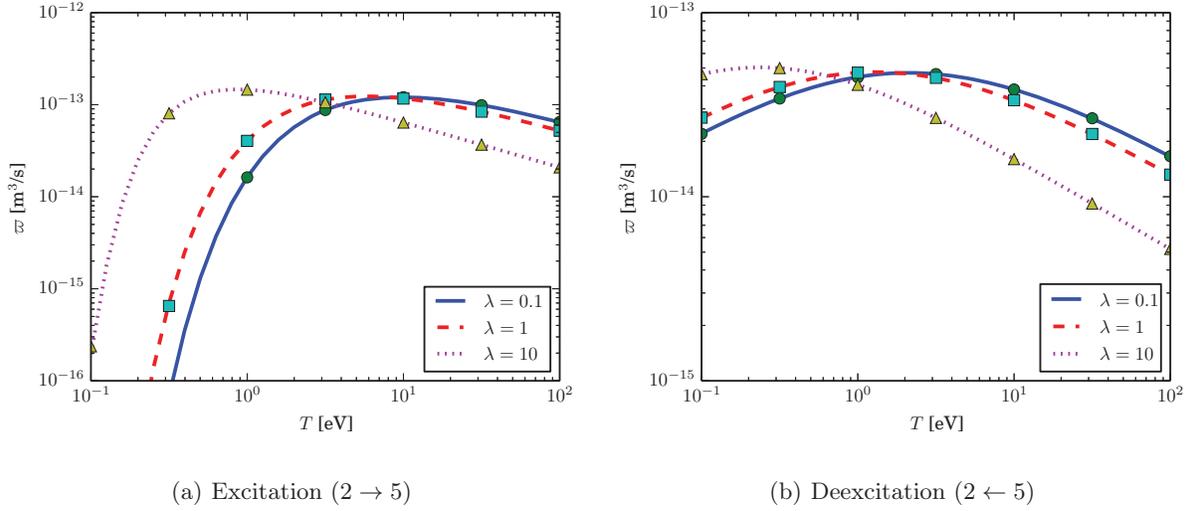}
\caption{Comparison of zeroth-order reaction rates with Monte Carlo integration of the full transfer integral. Lines are from our derived expressions. Symbols are Monte Carlo results. For each line, the ratio of the drift and thermal temperature $\lambda$ is fixed and the values are shown in the figure.}
\label{fig:MC}
\end{figure}

The accuracy of the derived formulas of the exchange source terms are verified against direct evaluation of the full transfer integral~(\ref{eq:ti3}) over six dimensional space using Monte Carlo method. The procedure for the Monte Carlo integration is as follows: (1) sample different pair of particles (one atom and one electron) from two different Maxwellian distributions, (2) compute the exchange rate due to each sample pair, and (3) accumulate these rates. The sum of these rates will follow the correct probability distribution function of the samples. For brevity, we only show an example calculation of the zeroth-order reaction rate for an excitation and deexcitation from levels $2$ and $5$. Figure~\ref{fig:MC} shows a detailed comparison between the rate expressions obtained from eqs. (\ref{eq:mf0_2a}) and (\ref{eq:mf0_2b}) against Monte Carlo results with an excellent agreement. Similar agreement is obtained with other sampled transitions, giving us complete confidence in the accuracy of our derivation of the multifluid rates from kinetic theory.

\subsection{Zero-dimensional Calculations}
\label{sec:numerix-4}

We conducted zero-dimensional (0D) calculations for a constant-volume (isochoric) system to support our findings in sections \ref{sec:numerix-1} and \ref{sec:numerix-2}, which include the following: (1) reaction rates can be enhanced when the relative mean drift velocity is significant, and (2) momentum transfer and thermal relaxation due to inelastic collisions are non-negligible. The system contains hydrogen atoms and a small fraction of free electrons and ions. Here we consider an electron-hydrogen two-fluid system. The relaxation between the heavy particles are assumed to be infinitely fast such that they can be described as a single heavy fluid with a bulk velocity $\bvec{u}_h$ and a temperature $T_h$. The governing equations are described in appendix \ref{app:G}, and the resultant system of ordinary differential equations is solved using the Radau5 method of Hairer and Wanner \cite{hairer_solving_1996}, which is ideally suited for stiff problems.

The initial conditions for the number densities and temperatures of the heavy particles and free electrons are summarized in Table \ref{tab:IC}. Initially, all the atoms are at the ground state (denoted by $k=1$ in table \ref{tab:IC}) with a translational temperature of $0.3$ eV. The atoms and ions are assumed to be at rest, i.e., their mean velocity is zero. The free electrons have a temperature of $2$ eV, and their mean velocity is varied to demonstrate the multifluid effects. While the electron number density is fixed for all the test cases shown here, the ion density is varied to examine the role of Coulomb collisions. In all the test cases, we include the first 10 atomic levels of hydrogen according to the model described in Appendix~\ref{app:F}. 

\begin{table}
\begin{tabular}{|c|c|c|}
\hline \hline
 & number density & temperature \\ 
\hline 
atomic & $n_k = 0.9n_t$ for $k=1$& 0.3 eV\\ 
 states    & $n_k = 10^{-15}n_t$ otherwise& \\ 
 \hline
  ion    & $n_i = f  n_e$; $f=0,0.2,1$ & 0.3 eV\\ 
\hline
electron & $n_e = 0.1n_t$ & 2 eV\\ 
\hline \hline
\end{tabular} 
\caption{Initial conditions of 0D test cases. The ion density is varied for each test case. For all cases, the total atomic density $n_t$ is $10^{20}$ m$^{-3}$. }
\label{tab:IC}
\end{table}

\begin{figure}
\centering
\includegraphics[scale=.85]{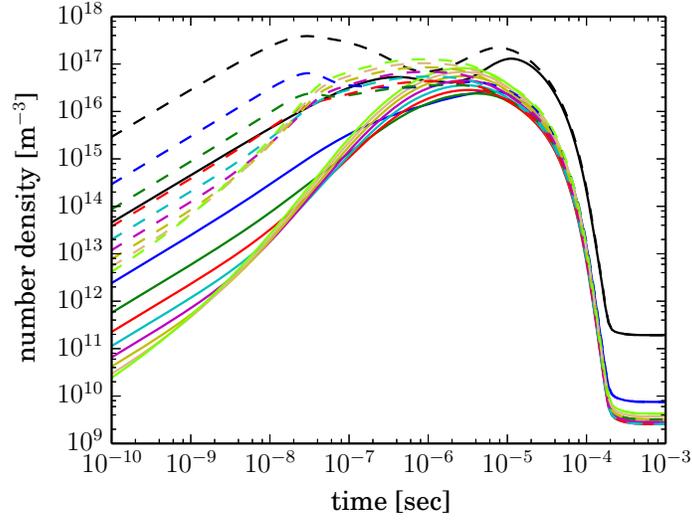}
\caption{Number density of excited states during a zero-dimensional chemistry test. Solid lines correspond to the solutions where the relative mean velocity is very small ($T_w=0.01$ eV). Dashed lines correspond to the solutions with a large relative mean velocity ($T_w = 10$ eV).}
\label{fig:Nevol}
\end{figure}

\begin{figure}
\centering
\includegraphics[scale=.85]{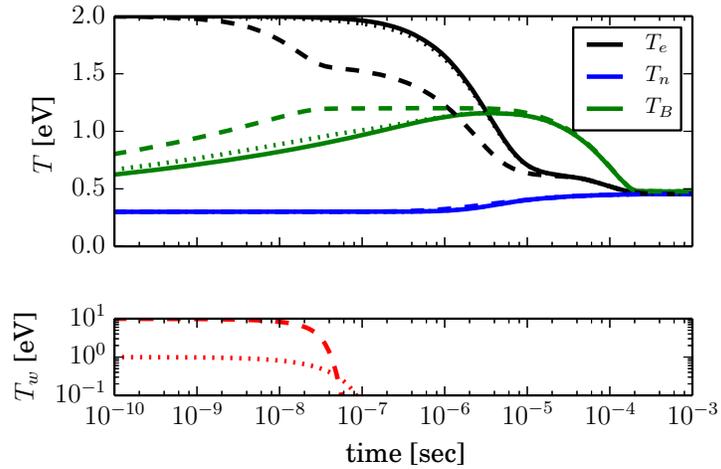}
\caption{Time evolution of the temperatures for several test cases with different initial drift velocities: solid line ($T_w=0.01$ eV), dotted line ($T_w=1$ eV), and dashed line ($T_w=10$ eV). The bottom plot shows the evolution of the drift temperature $T_w$.}
\label{fig:Tevol}
\end{figure}

In the first test, we perform the calculation with various initial mean velocities of the electrons, or equivalently, the drift temperatures $T_w$. The number density of ions is set to zero in this test, so there is no Coulomb collision. Figure \ref{fig:Nevol} shows the time evolution of the number densities of the excited states during the isochoric heating process for two extreme cases: $T_w = 0.01$ and $10$ eV; the former corresponds to a single-fluid calculation of a thermal bath with a warm electron population, and the latter corresponds to a situation where an electron beam is injected to the system. One can clearly see from Figure 
\ref{fig:Nevol} that there is an enhancement to the excitation process, indicated by an early increase in the population of excited states, due to the presence of a non-zero mean velocity of the electrons. The same argument can be made from Figure \ref{fig:Tevol}, which shows the time evolution of the temperatures for three cases of different initial $T_w$. In this plot, the Boltzmann temperature $T_B$ indicates the degree of excitation of the atom. It must be pointed out that the enhancement in excitation, however, persists on the time scale of the momentum relaxation process, which is indicated by a drop in $T_w$ at approximately $3 \times 10^{-7}$ sec as shown in the bottom plot of Figure \ref{fig:Tevol}. After this time, the momentum of the electrons is completely absorbed by the atom, signifying a change to single-fluid kinetics.
In all the test cases, the excitation proceeds at a time scale much smaller than the resolution of the figures, i.e., $T_B$ approximately goes from 0 to 0.7 eV in $10^{-9}$ sec. As mentioned before, when $T_B$ is sufficiently large, the momentum exchange and thermal relaxation rates from inelastic collisions cannot be neglected. 

\begin{figure}
\centering
\includegraphics[scale=.75]{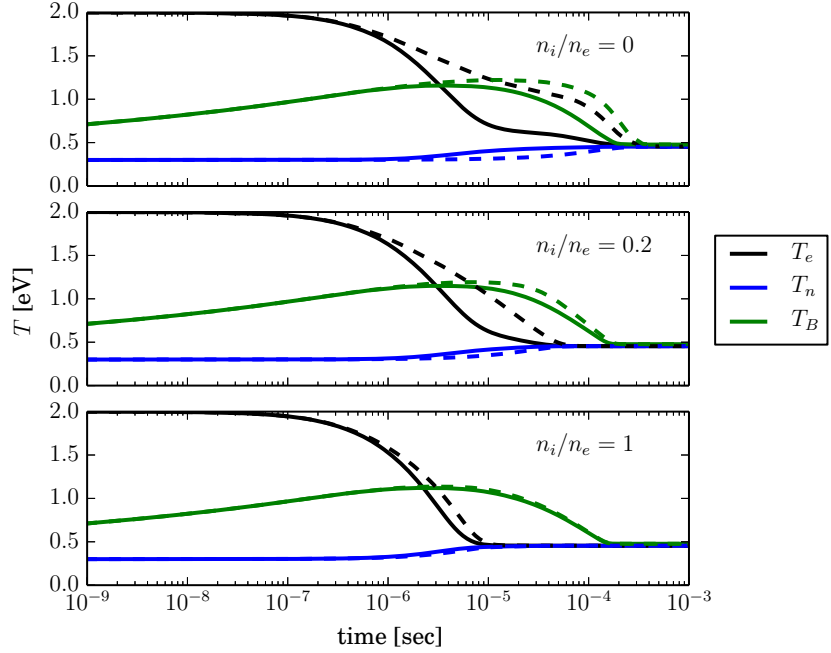}
\caption{Time evolution of the temperatures for the case with (solid) and without (dashed) the exchange terms (momentum and energy) due to inelastic collisions. The ion number densities are specified for each case. Initially, $T_w = 0.01$ eV.}
\label{fig:Tevol2}
\end{figure}

In the second test, we specifically identify the effect of inelastic collisions. Figure \ref{fig:Tevol2} shows a comparison of the temperature evolution for three test cases, all with a very small initial drift temperature $T_w = 0.01$: the solid lines correspond the solution with both elastic\footnote{These include Coulomb collisions in the case when $n_i \neq 0$.} and inelastic exchanges, and the dashed lines to the solution without inelastic exchanges. In these test cases, we slowly introduce the ions which effectively increase thermalization between the heavy particles and the electron via Coulomb collisions. The friction is negligible since $T_w \approx 0$, and there are only thermal relaxation and heat release from reaction. It is clearly shown that inelastic collisions do contribute to the total energy transfer between the electrons and atoms, leading to a faster temperature equilibration $T_h-T_e-T_B$. These effects are most noticeable when the ion density is low due to a weaker contribution of Coulomb collisions. Nevertheless, we observe under-prediction of temperature relaxation in all cases when inelastic energy exchanges are omitted. It is interesting to point out here that while $T_e-T_B$ equilibration is due to the heat release/absorption term, $T_e-T_h$ equilibration is only due to the thermal relaxation term. This result is quite intriguing, since the impact of inelastic collisions on thermal relaxation is normally neglected in most of single-fluid multi-temperature calculations.

\begin{figure}
\centering
\includegraphics[scale=.75]{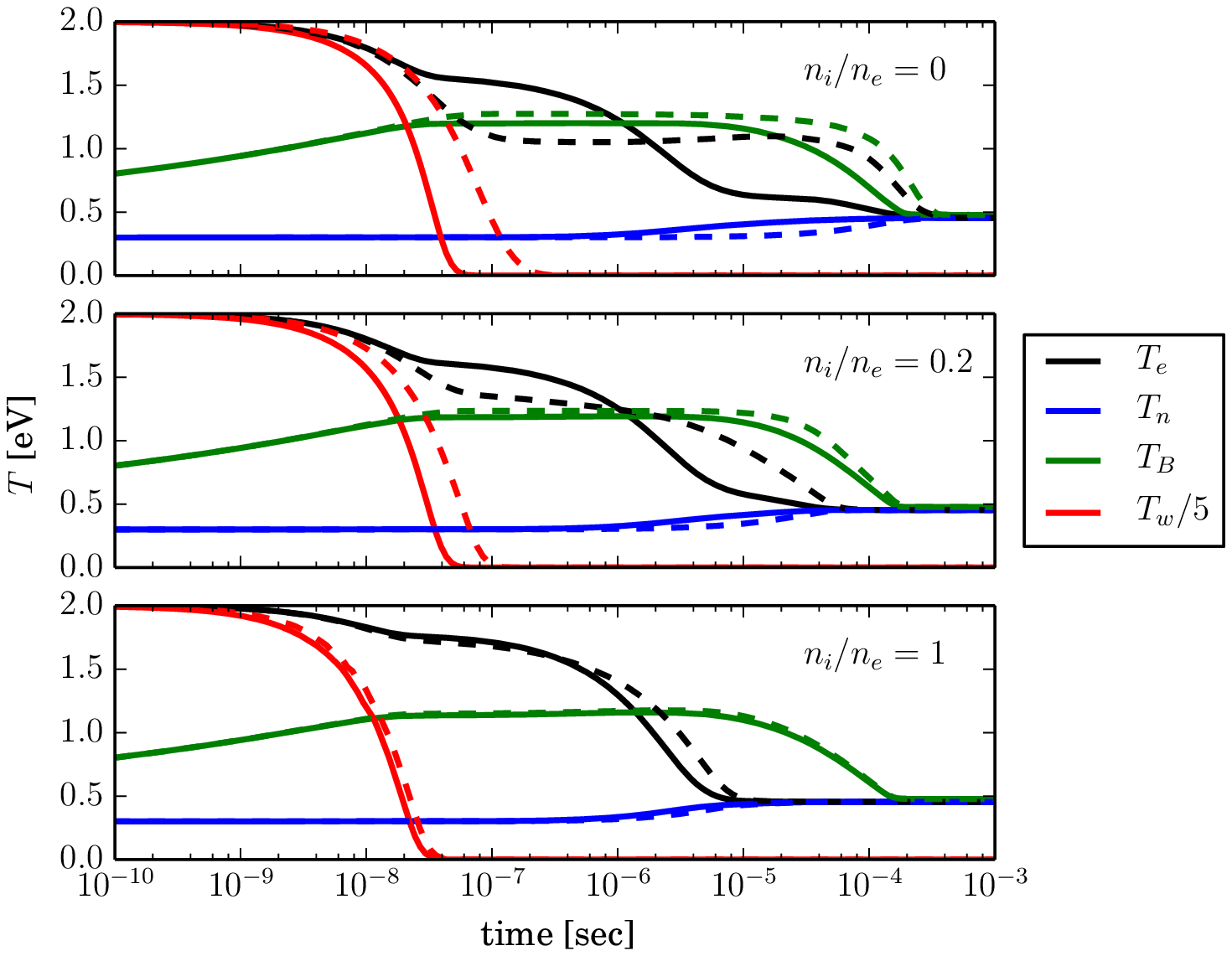}
\caption{Time evolution of the temperatures for the case with (solid) and without (dashed) the exchange terms (momentum and energy) due to inelastic collisions. The ion number densities are specified for each case. Initially, $T_w = 10$ eV.}
\label{fig:Tevol3}
\end{figure}

Figure \ref{fig:Tevol3} shows the results for the same three test cases above but now with an initial drift temperature $T_w = 10$ eV. One can make the same argument that inelastic collisions further enhance the temperature equilibration process. However, it is worthwhile to point out that during the time period $10^{-8} < t < 10^{-7}$, the electrons, although being decelerated due to friction, are also heated due to the work done by the same force (first term on the right hand side of eq. (\ref{eq:aG-3a})). This thermal heating term contains contribution from both elastic and inelastic collisions, which explains why the electron temperature drops faster towards equilibrium when inelastic exchanges are neglected. Finally we should emphasize that the thermal equilibrium between all components ($T_B,T_e,T_h$) can be obtained purely from inelastic collisions, as a result of properly accounting for detailed balance in our model; elastic collisions are not required to achieve equilibrium, but of course are needed to obtain the correct rate of relaxation.

\section{Concluding Remarks}
\label{sec:conclusion}

We have presented a model for inelastic collisions for electronic excitation and deexcitation within the context of a multifluid description of a plasma. The model is rigorously derived from kinetic theory and is applicable to any multifluid plasma, irrespective of the mass ratio, and strictly obeys the detailed balance principle. The appropriate mass transfer rates and momentum and thermal resistance coefficients are derived, and are found to satisfy the proper asymptotic limits, such that in the limit of vanishing energy gap, the well-known expressions for elastic collisions are recovered. 

Numerical evaluations of the multifluid rates are carried out for a two-fluid electron-hydrogen plasma using Bohr model for the energy levels and semi-classical cross sections. Several numerical tests were performed in a virtual (zero-dimensional) test cell, and both the known thermal and beam limits were correctly recovered from the model. We also found that in some plasma conditions of interest, the contribution of the inelastic collisions to the resistance coefficients is significant, contrary to the usual assumptions made in current multifluid models. While this work is focused on two-body reactions, extension to ionization and recombination processes is currently under examination, and will be presented in a future article.

\section*{Acknowledgments}
The authors thank Drs. Ann Karagozian, Justin Koo, Robert Martin, and Owen Smith for stimulating discussions. Research was supported by the Air Force Office of Scientific Research (AFOSR), grant numbers 12RZ06COR (PM: Dr. F. Fahroo) and 14RQ05COR (PM: Dr. J. Marshall).

\bibliography{refs}

\numberwithin{equation}{section}
\begin{appendices}
\section{Collision kinematics}
\label{app:A}

Let us consider an inelastic collision between two particles $s$ and $t$, such that the post-collision particles can have modified internal states. The process is formally described as the relation
\begin{align}\label{eq:a1}
s(\bvec{v}_s) + t(\bvec{v}_t) \rightarrow s'(\bvec{v'}_s) + t'(\bvec{v'}_t)
\end{align} 
Note that only two particles are produced by the collision.
The initial velocities are $\bvec{v}_s, \bvec{v}_t$. The mean fluid velocity is $\bvec{u}$, such that $\bvec{u} \equiv \langle \bvec{v} \rangle \equiv \int d^3 \bvec{v} \bvec{v} f(\bvec{v})$ and a thermal velocity $\bvec{c} = \bvec{v} - \bvec{u}$. By definition, we also have $\langle \bvec{c} \rangle \equiv 0$.

The collision can be transformed to the center of mass (COM) reference frame, moving with velocity $\mathbf{V}$ with respect to the LAB frame. 
Similarly, we can also define a mean velocity of this COM frame as $\bvec{U}$. The subsequent Galilean transformations yield the following definitions:
\begin{subequations}\label{eq:a2a}
\begin{align} 
	\bvec{V} = \frac{m_s\bvec{v}_s +m_t\bvec{v}_t}{M}\qquad  & \bvec{g} = \bvec{v}_s-\bvec{v}_t\\
	\bvec{U} = \frac{m_s\bvec{u}_s +m_t\bvec{u}_t}{M}\qquad & \bvec{w} = \bvec{u}_s-\bvec{u}_t
\end{align}
\end{subequations}
where $M\!=\!m_s\!+\!m_t$.
The inverse transformation yields:
\begin{subequations}\label{eq:a2b}
\begin{align} 
	\bvec{v}_s = \bvec{V} +\frac{m_t}{M}\bvec{g} \qquad  & \bvec{u}_s = \bvec{U} +\frac{m_t}{M}\bvec{w}\\
	\bvec{v}_t = \bvec{V} -\frac{m_s}{M}\bvec{g}\qquad    & \bvec{u}_t = \bvec{U} -\frac{m_s}{M}\bvec{w}
\end{align}
\end{subequations}
Mass conservation imposes the relation $m_s\!+\!m_t \!=\! M \!=\! m_s'\!+\!m_t'$. For the case of two-body processes such as excitation of internal states, the masses are individually conserved, i.e. $m_s' \!=\! m_s, m_t'\!=\! m_t$.
Expressed in the COM frame, momentum and energy conservation yield, respectively:
\begin{subequations}\label{eq:a3}
\begin{align}
M\bvec{V}= &M\bvec{V'}\\
\frac{1}{2}M\bvec{V}^2 + \frac{1}{2}\mu\bvec{g}^2 =& \frac{1}{2}M\bvec{V'}^2 + \frac{1}{2}\mu\bvec{g'}^2 + \Delta\varepsilon
\end{align}
\end{subequations}
where $\mu\!=\!m_sm_t/M$. Therefore, we have the following constraints:
\begin{equation}\label{eq:a4}
\bvec{V} = \bvec{V'}\qquad\textrm{and}\quad \bvec{g}^2=\bvec{g'}^2+\frac{2\Delta\varepsilon}{\mu}
\end{equation}
For an excitation between two atomic levels, the transferred energy is a fixed value $\Delta\varepsilon\equiv \varepsilon^* > 0$, the energy gap between levels. For a deexcitation, we use the same formulation as above (i.e. post-collision variables indicated by a prime), but in this case, $\Delta\varepsilon = -\varepsilon^* < 0$. In the limit $\Delta\varepsilon\rightarrow 0$, the collision is elastic.

\section{Separation of variables}
\label{app:B}

Consider the Maxwellian velocity distribution functions (VDF) of each particle type, normalized to unity, e.g. (recall that $\bvec{c}\!=\!\bvec{v}\!-\!\bvec{u}$):
\begin{equation}\label{eq:b1}
f_s(\bvec{v}_s) = \left(\frac{m_s}{2\pi kT_s}\right)^{\frac{3}{2}} \exp\left[-\frac{m_s\bvec{c}_s^2}{2kT_s}\right]
\end{equation}
and similarly for $f_t$. The averaging over initial states will yield a product of these two distributions:
\begin{equation}\label{eq:b2}
f_s(\bvec{v}_s)f_t(\bvec{v}_t) = \left(\frac{m_s}{2\pi kT_s}\right)^{\frac{3}{2}} \left(\frac{m_t}{2\pi kT_t}\right)^{\frac{3}{2}} \exp[\mathcal{A}]
\end{equation}
where the argument of the exponential function is, from (\ref{eq:a2b}):
\begin{equation}\label{eq:b3}
\mathcal{A} = \frac{m_s}{2 kT_s}\left[\bvec{V}-\bvec{U}+\frac{m_t}{M}(\bvec{g}-\bvec{w})\right]^2 + \frac{m_t}{2 kT_t}\left[\bvec{V}-\bvec{U}-\frac{m_s}{M}(\bvec{g}-\bvec{w})\right]^2
\end{equation}
Following Burgers~\cite{burgers_flow_1969}, this expression can be simplified with an appropriate transformation of variables; since the basic procedure will be used elsewhere, we describe it below.
First, we define the following variables
\begin{equation}\label{eq:b4}
\beta_p = \frac{m_p}{2kT_p},\qquad \tvec{g}=\bvec{g}-\bvec{w}
\end{equation}
such that
\begin{align}\label{eq:b5}
\mathcal{A} &= \beta_s\left[(\bvec{V}\!-\!\bvec{U})+\frac{m_t}{M}\tvec{g}\right]^2 +\beta_t\left[(\bvec{V}\!-\!\bvec{U})-\frac{m_s}{M}\tvec{g}\right]^2\nonumber\\
 &=(\beta_s\!+\!\beta_t)(\bvec{V}\!-\!\bvec{U})^2 +\left[\beta_s\frac{m_t^2}{M^2}+\beta_t\frac{m_s^2}{M^2}\right] \tvec{g}^2 
    + 2\left[\beta_s\frac{m_t}{M}-\beta_t\frac{m_s}{M}\right] (\bvec{V}\!-\!\bvec{U})\cdot\tvec{g}
\end{align}
Let us define:
\begin{equation}\label{eq:b6}
\bvec{V^*}=\bvec{V}\!-\!\bvec{U} + \gamma \tvec{g}
\end{equation}
and comparing the expression
\begin{equation}\label{eq:b7}
(\beta_s\!+\!\beta_t)\bvec{V^*}^2 = (\beta_s\!+\!\beta_t)(\bvec{V}\!-\!\bvec{U})^2 +(\beta_s\!+\!\beta_t)\gamma^2\tvec{g}^2 +2\gamma(\beta_s\!+\!\beta_t)(\bvec{V}\!-\!\bvec{U})\cdot\tvec{g}
\end{equation}
with~(\ref{eq:b5}), we can choose the appropriate value of the coefficient $\gamma$ to eliminate the dot product from $\mathcal{A}$:
\begin{equation}\label{eq:b8}
\gamma = \frac{1}{\beta_s\!+\!\beta_t}\left(\beta_s\frac{m_t}{M}-\beta_t\frac{m_s}{M}\right)
\end{equation}
We then obtain complete separation of variables:
\begin{equation}\label{eq:b9}
\mathcal{A} = ({\beta_s\!+\!\beta_t})\bvec{V^*}^2 +\left[ \beta_s\frac{m_t^2}{M^2}\!+\!\beta_t\frac{m_s^2}{M^2}-\frac{1}{\beta_s\!+\!\beta_t} \left(\beta_s\frac{m_t}{M}-\beta_t\frac{m_s}{M}\right)^2\right] \tvec{g}^2
\end{equation}
The term in brackets is easily simplified:
\begin{equation}\label{eq:b10}
\left[\ldots\right] = \frac{\beta_s\beta_t}{\beta_s+\beta_t}
\end{equation}
We can now define effective, average temperatures:
\begin{subequations}\label{eq:b11}
\begin{align}
\beta_s \!+\!\beta_t &= \frac{m_s}{2kT_s}\!+\!\frac{m_t}{2kT_t}=\frac{M}{2k}\frac{m_sT_t+m_tT_s}{MT_sT_t}\equiv \frac{M}{2kT^*}\\
\frac{\beta_s\beta_t}{\beta_s+\beta_t} &= \frac{\mu}{2k}\,\frac{M}{T_sT_t}\,\frac{T_sT_t}{m_sT_t\!+\!m_tT_s}\equiv \frac{\mu}{2k\tilde{T}}
\end{align}
\end{subequations}
and $\gamma$ becomes:
\begin{equation}\label{eq:b12}
\gamma = \frac{\mu}{M} \, \frac{T_t-T_s}{\tilde{T}} = \mu\,\frac{T_t-T_s}{m_sT_t+m_tT_s}
\end{equation}
To summarize, we have performed the following change of variables:
\begin{subequations}\label{eq:b13}
\begin{align}
	\bvec{V^*} &= \bvec{V}-\bvec{U}+\mu \frac{T_t-T_s}{m_sT_t\!+\!m_tT_s}\tvec{g} \quad & \tvec{g} &=\bvec{g}-\bvec{w}\quad &{}\\
	T^*&= M\frac{T_sT_t}{m_sT_t\!+\!m_tT_s}\quad  & \tilde{T}&=\frac{m_sT_t\!+\!m_tT_s}{M}\quad &{}
\end{align}
\end{subequations}
These are the same expressions found in~\cite[pp.~45-46]{burgers_flow_1969} (with an occasional change of naming convention)
for which it is easy to verify that the Jacobian of the transformations is unity, i.e.
\begin{equation}\label{eq:b14}
	d^3\bvec{v}_sd^3\bvec{v}_t \equiv d^3\bvec{V}d^3\bvec{g} \equiv d^3\bvec{V^*}d^3\tvec{g}
\end{equation}
Furthermore, we note that:
\begin{equation}\label{eq:b15}
\left(\frac{m_s}{2kT_s}\right)^{\frac{3}{2}} \left(\frac{m_t}{2kT_t}\right)^{\frac{3}{2}} \equiv \left(\beta_s\beta_t\right)=(\beta_s\!+\!\beta_t)^{\frac{3}{2}}\left(\frac{\beta_s\beta_t}{\beta_s\!+\!\beta_t}\right)^{\frac{3}{2}}\equiv \left(\frac{M}{2kT^*}\right)^{\frac{3}{2}} \left(\frac{\mu}{2k\tilde{T}}\right)^{\frac{3}{2}} 
\end{equation}

The product of two distributions can now be written as:
\begin{equation}\label{eq:b16}
f_s\cdot f_t = \left(\frac{M}{2\pi kT^*}\right)^{\frac{3}{2}} \!\exp\left[-\frac{M\bvec{V}^{*2}}{2kT^*}\right] \cdot  \left(\frac{\mu}{2\pi k\tilde{T}}\right)^{\frac{3}{2}} \!\exp\left[-\frac{\mu\tvec{g}^{2}}{2k\tilde{T}}\right] \equiv f^*(\bvec{V^*})\cdot \tilde{f}(\tvec{g})
\end{equation}
All subsequent expressions can now be simplified with this separation of variables. For example, any operator $\mathscr{O}$ that depends only on variables expressed in the COM frame ($\bvec{g},\bvec{g'}$), we have:
\begin{equation}\label{eq:b17}
\int d^3\bvec{v}_s d^3\bvec{v}_t f_s f_t \mathscr{O}(\bvec{g},\bvec{g}')= \underbrace{\int d^3\bvec{V}^* f^*(\bvec{V}^*)}_{\equiv 1} \cdot \int d^3\tvec{g} \tilde{f}(\tvec{g}) \mathscr{O}(\bvec{g},\bvec{g}')
\end{equation}

Note that this procedure applies equally well for elastic, excitation and deexcitation collisions, and that no approximations have been made on the mass ratio. Furthermore, since the averaging over initial states only involves the distribution functions for the $s$ and $t$ particles, we have not necessarily assumed that the final products $s'$ and $t'$ belong to the same fluid as the initial particles.

\section{Atomic data and cross section models}
\label{app:F} 
The atomic states of the Hydrogen atom are listed as a function of their principal quantum number ($n$) only, following the Bohr atomic model; the splitting of states with respect to orbital and spin numbers is ignored, and all states have a degeneracy $\mathpzc{g}_n=2n^2$. The states number from $n=1$ to $\infty$ and we consider a finite number of states $n=1,\ldots,M < \infty$ before reaching the ionization limit\footnote{Strictly speaking, the ionization limit $I_H$ is attained for $n\rightarrow\infty$. In reality, the ionization potential is lowered as a result of interaction with the plasma (Debye shielding) and quantum uncertainty. In practice, the truncation is accomplished at a lower limit still; for the current purpose, details of this truncation procedure can be ignored.  Suffice to say that the series extends to a number $n=M$, which can be considered large, e.g. $O(100)$.}.
In this simplified model, the energy of each state is given as $E_n\!=\! I_H\left(1\!-\!1/n^2\right)$, as measured from the ground state  ($E_1\equiv 0$), and we will denote by $I_n\!=\!I_H \left(1/n^2\!-\!1/M^2\right)\!\simeq\! I_H/n^2$ the energy required for ionization of level $n$.

The classical form of the cross section for energy exchange between a free electron and the atom (Hydrogen) is used \cite{zeldovich_physics_2002}. For an excitation collision from level $\ell$ to level $u>\ell$, the cross section takes the form:
\begin{align}
\sigma^{\uparrow}_{(u|\ell)} (x) = (4\pi a_o^2) (3\textrm{f}_{\ell u}) \left( \frac{I_H}{kT_e} \right)^2 \frac{ (x - x_{\ell u})}{x_{\ell u} x^2}
\end{align}
where $a_o$ is the Bohr radius, $x$ is the nondimensional incident energy of the electron, $x_{\ell u} = (E_u - E_\ell)/kT_e$ is the energy gap between $\ell$ and $u$, and $\textrm{f}_{\ell u}$ is the oscillator strength:
\begin{equation}
\textrm{f}_{\ell u} = \frac{32}{3\pi\sqrt{3}}\frac{1}{\ell^5}\frac{1}{u^3} \frac{1}{\left(\frac{1}{\ell^2}-\frac{1}{u^2}\right)^3}
\end{equation}

In the thermal (single-fluid) limit ($\lambda \rightarrow 0$), the reaction rate can be obtained in an exact form:
\begin{align}\label{eq:AF-1}
\varpi^{\uparrow}_{(u|\ell)} \simeq (4\pi a_o^2) (3\textrm{f}_{\ell u}) \bar{v}_e \left( \frac{I_H}{kT_e} \right)^2 \psi_{\ell u}
\end{align}
where
\begin{equation}
\bar{v}_e =  \left(\frac{8kT_e}{\pi m_e}\right)^\frac{1}{2},\ \ \ \psi_{\ell u} = \frac{e^{-x_{\ell u}}}{x_{\ell u}} - \textrm{E}_1(x_{\ell u}) \qquad\textrm{and}\qquad \textrm{E}_1(x) \!=\! \int_x^\infty\! \frac{e^{-y}}{y}dy
\end{equation}
Here, $\bar{v}_e$ is the mean thermal electron velocity and $\textrm{E}_1$ is the exponential integral. On the other hand, in the beam limit ($T_e \rightarrow 0$), the reaction rate takes the form:
\begin{align}\label{eq:AF-2}
\varpi^{\uparrow}_{(u|\ell)} \simeq \frac{\sqrt{\pi}}{2} \bar{v}_e \sigma^\uparrow_{(u|\ell)} (\lambda) = (4\pi a_o^2) (3\textrm{f}_{\ell u}) \bar{v}_e \left( \frac{I_H}{kT_e} \right)^2 \phi_{\ell u} 
\end{align}
where
\begin{align}
\phi_{\ell u} = \frac{\sqrt{\pi}}{2} \frac{ (\lambda - x_{\ell u})}{x_{\ell u} \lambda^2}
\end{align}

\section{Rate equations}
\label{app:G} 
We describe here the governing equations for the zero-dimensional simulation, which describe the time evolution of a constant-volume system during a thermochemical relaxation process. Here we made the approximation that $m_n \simeq m_i \simeq m_h$. The following system of rate equations is considered:
\begin{subequations}
\begin{align}
\frac{d n_e}{dt} &= 0 \label{eq:aG-1a} \\
\frac{d n_i}{dt} &= 0 \\
\frac{d n_k}{dt} &= \Gamma_k^{(xd)} \label{eq:aG-1b} \\
\frac{d }{dt}(m_e n_e \mathbf{u}_e) &= - (\mathbf{u}_e-\mathbf{u}_h) (K^{(en)} + K^{(ei)} + K^{(xd)}) \label{eq:aG-1c} \\
\frac{d }{dt}(m_h n_h \mathbf{u}_h)  &= + (\mathbf{u}_e-\mathbf{u}_h) (K^{(en)} + K^{(ei)} + K^{(xd)}) \label{eq:aG-1d} \\
\frac{d E_e}{dt} &= - \mathbf{\bar{u}} \cdot (\mathbf{u}_e-\mathbf{u}_h) (K^{(en)} + K^{(ei)} + K^{(xd)}) \nonumber \\
&- \frac{2 m_e}{m_h} k(T_e-T_h) (J^{(en)} + J^{(ei)} + J^{(xd)}) - \Xi^{(xd)} \label{eq:aG-1e} \\
\frac{d E_h}{dt} &= + \mathbf{\bar{u}} \cdot (\mathbf{u}_e-\mathbf{u}_h) (K^{(en)} + K^{(ei)} + K^{(xd)}) \nonumber \\
&+ \frac{2 m_e}{m_h} k(T_e-T_h) (J^{(en)} + J^{(ei)} + J^{(xd)})  \label{eq:aG-1f} 
\end{align}
\end{subequations}
where $n_h = n_i + \sum_{k \in n} n_k$ and $E_{e(h)}$ is the total energy of the fluid:
\begin{align}\label{eq:aG-2}
E_{e(h)} &= \underbrace{ \frac{3}{2} n_{e(h)} k T_{e(h)} }_{\varepsilon_{e(h)}} + \frac{1}{2} m_{e(h)} n_{e(h)} \mathbf{u}_{e(h)} \cdot \mathbf{u}_{e(h)}
\end{align}
where $\varepsilon_{e(h)}$ is defined as the thermal energy of the fluid. The exchange source terms can be decomposed into three parts: elastic and inelastic collisions between the electrons and atoms and Coulomb collisions between the charged particles. The momentum and energy exchange rates for these processes are defined in eqs.~(\ref{eq:n-1}), (\ref{eq:kCoul}) and (\ref{eq:n-2}). The remaining terms are:
\begin{align}
\Gamma_k^{(xd)} &= \sum_{u > k} \left[ -n_k n_e \varpi^{\uparrow}_{(u|k)} +n_u n_e \varpi^{\downarrow}_{(k|u)} \right] 
+ \sum_{\ell < k} \left[ -n_k n_e \varpi^{\downarrow}_{(\ell|k)} +n_\ell n_e \varpi^{\uparrow}_{(k|\ell)} \right]\\
\Xi^{(xd)} &= \sum_\ell \sum_{u>\ell} \left[ n_\ell n_e \varpi^{\uparrow}_{(u|\ell)} \varepsilon^*_{\ell u} - n_u n_e \varpi^{\downarrow}_{(\ell|u)} \varepsilon^*_{\ell u}\right]
\end{align}
where $\varepsilon^*_{\ell u} = E_u - E_\ell$. Comparing eqs. (\ref{eq:aG-1c})-(\ref{eq:aG-1f}) and (\ref{eq:aG-2}), we can also write conservation equations for the thermal energies of electrons and atoms:
\begin{subequations}
\begin{align}
\frac{d \varepsilon_e}{dt} &= - m_e (\mathbf{\bar{u}}- \mathbf{u}_e)\cdot (\mathbf{u}_e-\mathbf{u}_h) (K^{(en)} + K^{(ei)} + K^{(xd)}) \nonumber \label{eq:aG-3a}\\
&- \frac{2 m_e}{m_h} k(T_e-T_h) (J^{(en)} + J^{(ei)} + J^{(xd)}) - \Xi^{(xd)}\\
\frac{d \varepsilon_h}{dt} &= + m_e (\mathbf{\bar{u}} - \mathbf{u}_h) \cdot (\mathbf{u}_e-\mathbf{u}_h) (K^{(en)} + K^{(ei)} + K^{(xd)}) \nonumber \\
&+ \frac{2 m_e}{m_h} k(T_e-T_h) (J^{(en)} + J^{(ei)} + J^{(xd)})  \label{eq:aG-3b}
\end{align}
\end{subequations}

\end{appendices}

\end{document}